\documentstyle[11pt,aaspp4,flushrt]{article}
\slugcomment{\it Invited Review, to appear in 1999 December PASP}
\lefthead{Stern \& Spinrad}
\righthead{Search Techniques for Distant Galaxies}

\def\cf{{c.f.,~}}
\def\ie{{i.e.,~}}
\def\eg{{e.g.,~}}
\def\etal{{et al.~}}

\def\deg{\ifmmode {^{\circ}}\else {$^\circ$}\fi}

\def\secper{\ifmmode \rlap.{^{s}}\else $\rlap{.}{^{s}} $\fi}

\def\kms{\ifmmode {\rm\,km\,s^{-1}}\else
    ${\rm\,km\,s^{-1}}$\fi}
\def\kmsMpc{\ifmmode {\rm\,km\,s^{-1}\,Mpc^{-1}}\else
    ${\rm\,km\,s^{-1}\,Mpc^{-1}}$\fi}
\def\ergAcm2{\ifmmode {\rm\,ergs\,cm^{-2}\,{\rm \AA}^{-1}}\else
    ${\rm\,ergs\,cm^{-2}\,\AA^{-1}}$\fi}
\def\ergcm2s{\ifmmode {\rm\,ergs\,cm^{-2}\,s^{-1}}\else
    ${\rm\,ergs\,cm^{-2}\,s^{-1}}$\fi}
\def\ergsHz{\ifmmode {\rm\,ergs\,s^{-1}\,Hz^{-1}}\else
    ${\rm\,ergs\,s^{-1}\,Hz^{-1}}$\fi}
\def\ergs{\ifmmode {\rm\,ergs\,s^{-1}}\else
    ${\rm\,ergs\,s^{-1}}$\fi}
\def\ergsA{\ifmmode {\rm\,ergs\,s^{-1}\,\AA^{-1}}\else
    ${\rm\,ergs\,s^{-1}\,\AA^{-1}}$\fi}
\def\WHz{\ifmmode {\rm\,W\,Hz^{-1}}\else
    ${\rm\,W\,Hz^{-1}}$\fi}
\def\WHzsr{\ifmmode {\rm\,W\,Hz^{-1}\,sr^{-1}}\else
    ${\rm\,W\,Hz^{-1}\,sr^{-1}}$\fi}
\def\ergscmHz{\ifmmode {\rm\,ergs\,cm^{-2}\,Hz^{-1}}\else
    ${\rm\,ergs\,cm^{-2}\,Hz^{-1}}$\fi}

\def\spose#1{\hbox to 0pt{#1\hss}}
\def\simlt{\mathrel{\spose{\lower 3pt\hbox{$\mathchar"218$}}
     \raise 2.0pt\hbox{$\mathchar"13C$}}}
\def\simgt{\mathrel{\spose{\lower 3pt\hbox{$\mathchar"218$}}
     \raise 2.0pt\hbox{$\mathchar"13E$}}}

\def\lya{Ly$\alpha$}

\begin{document}

\title{Search Techniques for Distant Galaxies}

\author{Daniel Stern \& Hyron Spinrad }
\affil{Department of Astronomy, University of California at Berkeley \\
Berkeley, CA 94720 \\
{\tt email: (dan,spinrad)@bigz.berkeley.edu}}

\begin{abstract}

How and when do galaxies form?  Studies of the microwave background
radiation reveal that the Universe is spectacularly homogenous at
redshift $z \approx 1000$.  Observations of the local Universe reveal
that by $z = 0$ much of the luminous matter has condensed into mature,
gravitationally-bound structures.  One of the primary challenges to
astronomers today is to achieve a robust understanding of this process
of galaxy formation and evolution.  Locating and studying young galaxies
at large look-back times is an essential aspect of this program.

We review the systematic observational techniques used to identify
galaxies at early cosmic epochs.  In the past few years, the study of
normal, star-forming galaxies at $z > 3$ has become possible; indeed,
successful methods have been developed to push the frontier past $z =
5$.  We are now directly observing individual galaxies within a Gyr of
the Big Bang.  We present a detailed review of the many search methods
used for identifying distant galaxies, consider the biases inherent in
different search strategies, and discuss early results of these
studies.  We conclude with goals for future studies at the start of the
21{\it st} century.

\end{abstract}

\section{Introduction}

At redshift $z \approx 1000$, the distribution of matter in the
Universe was remarkably smooth:  density fluctuations in the cosmic
microwave background were of order one part in $10^{5}$ on the degree
scale \markcite{Bennett:96}(\eg Bennett {et~al.} 1996).  Locally, $13~ h_{50}^{-1}~$ Gyr later
at $z = 0$, we observe that the distribution of baryonic matter on the
Mpc-scale is far from smooth, with baryons largely consigned to
luminous, bound structures, such as galaxies and clusters of
galaxies.   These present-day structures can be explained by the
gravitational collapse and coalescence of the overdense regions of the
early Universe.  A detailed understanding of this collapse, identified
as galaxy and large-scale structure formation, is uncertain currently,
and stands as one of the primary challenges to astrophysicists today.

The earliest epoch of galaxy formation lies beyond a redshift of 5.
Recent observations have, for the first time, directly measured systems
at the large lookback times implied by $z > 5$ \markcite{Dey:98,
Weymann:98, Spinrad:98, Chen:99, vanBreugel:99a, Hu:99}(\eg Dey {et~al.} 1998; Weymann {et~al.} 1998; Spinrad {et~al.} 1998; Chen, Lanzetta, \& Pascarelle 1999; van Breugel {et~al.} 1999; Hu, McMahon, \& Cowie 1999).  Several lines
of evidence support a substantial epoch of galaxy formation prior to $z
= 5$: the presence of metals (in excess of the primordial abundances)
in high-$z$ damped \lya\ systems \markcite{Lu:96}(\eg Lu {et~al.} 1996), quasars
\markcite{Hamann:99}(Hamann \& Ferland 1999), and star-forming galaxies at $z \sim 2.5 - 3.5$
\markcite{Steidel:96a, Lowenthal:97}(Steidel {et~al.} 1996a; Lowenthal {et~al.} 1997) requires metal creation and dispersal
at higher redshifts.  The tight photometric sequences in both low-$z$
and intermediate-$z$ clusters also attests to high formation redshift
$z_f$ at least for the elliptical galaxy formation in dense
environments \markcite{Stanford:95}(\eg Stanford, Eisenhardt, \&  Dickinson 1995).  Indeed, some ellipticals at
$z \sim 1.5$ are observed to contain evolved stellar populations with
ages in excess of 3.5 Gyr \markcite{Dunlop:96, Spinrad:97,
Dey:99b}(\eg Dunlop {et~al.} 1996; Spinrad {et~al.} 1997; Dey {et~al.} 2000), again implying high formation redshifts.

Theoretical paradigms of galaxy formation are vastly different:  do
large galactic spheroids form primarily via the monolithic collapse of
a protogalactic cloud \markcite{Eggen:62}(\eg Eggen, Lynden-Bell, \& Sandage 1962) or are they built up
through the hierarchical accretion of a multitude of subgalactic clumps
\markcite{Baron:87, Baugh:98}(\eg Baron \& White 1987; Baugh {et~al.} 1998)?  Both faint number counts and the
apparent lack of massive red systems at $z \simgt 1$ in ($K$-selected)
redshift surveys would seem to favor the latter model
\markcite{Kauffmann:98}(Kauffmann \& Charlot 1998).  However, the most direct answer will come with
detailed studies of protogalaxies in the early Universe.

\begin{figure}[!ht]
\plotfiddle{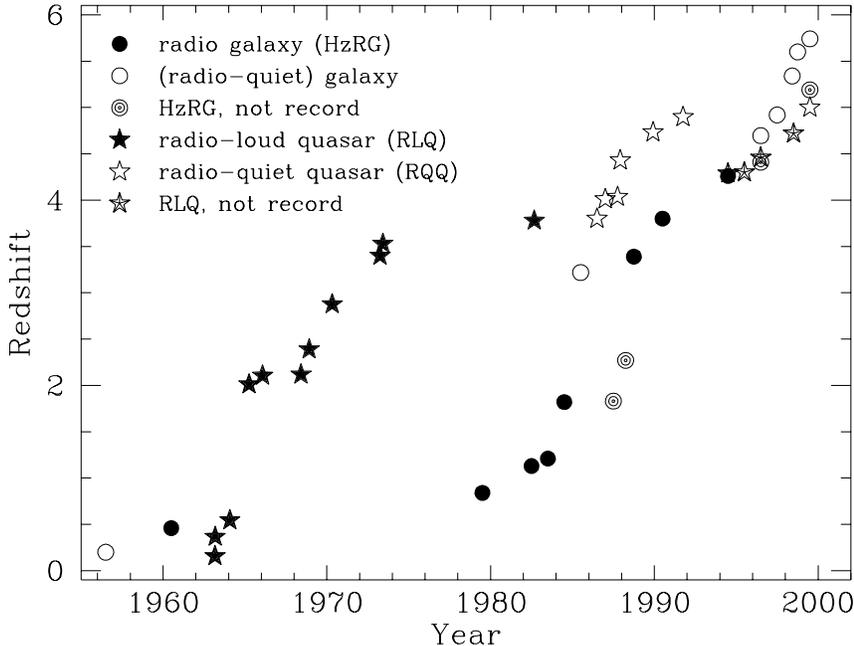}{3.2in}{-90}{45}{45}{-185}{255}

\caption[The highest-redshift object as a function of time]{The
highest-redshift object as a function of time.  Note that the
most distant quasars and galaxies were primarily selected from
radio emission until recent years (1985 for quasars, 1997 for
galaxies).}

\label{goldfig}
\end{figure}

Considerable astronomical expertise and resources have been levied at
identifying protogalaxies in the early Universe over the past 40 years
\markcite{Pritchet:94}(for a recent review, see Pritchet 1994).  Table~\ref{gold}
lists the most distant galaxy confirmed as a function of time (see
Fig.~\ref{goldfig}).  There are several established and innovative
methods to locate the minority population of distant systems from the
confusion of faint, intermediate-luminosity systems that dominate faint
galaxy counts (at optical/near-infrared wavelengths).  This manuscript
presents a review of these techniques with some attention applied to
the implications of the current studies and expectations for this line
of research in the near future.

% FIGURE 1

% TABLE 1
\begin{table}[ht!]
\caption{The Highest-Redshift Galaxy}
\footnotesize
\begin{center}
\begin{tabular}{llcll}
\hline\hline
Month/Year &
Galaxy &
$z$ &
Search Technique &
Reference \\
\hline
1999 & SSA22$-$HCM1 & 5.74 & narrow-band imaging & \markcite{Hu:99}Hu {et~al.} (1999) \\
1998 Oct & HDF~4-473.0  & 5.60 & photometric selection & \markcite{Weymann:98}Weymann {et~al.} (1998) \\
1998 May & 0140+326 RD1 & 5.34 & serendipity & \markcite{Dey:98}Dey {et~al.} (1998) \\
1997 & Cl~1358+62, G1/G2 arcs & 4.92 & serendipity/grav. lensing & \markcite{Franx:97}Franx {et~al.} (1997) \\
1996 & BR~1202$-$0725 companion & 4.70 & narrow-band imaging & \markcite{Petitjean:96}Petitjean {et~al.} (1996) \\
%1996 & 6C~0140+326  & 4.41 & radio selection & \markcite{Rawlings:96}Rawlings {et~al.} (1996) -- HzRG record \\
1994 & 8C~1435+63   & 4.26 & radio selection & \markcite{Lacy:94}Lacy {et~al.} (1994) \\
1990 & 4C~41.17     & 3.80 & radio selection & \markcite{Chambers:90}Chambers, Miley, \& van Breugel (1990) \\
1988 & B2~0902+34 & 3.39 & radio selection & \markcite{Lilly:88}Lilly (1988) \\
1985 & PKS~1614+051 companion & 3.215 & narrow-band imaging & \markcite{Djorgovski:85}Djorgovski {et~al.} (1985) \\
%1988 Apr & 4C~40.36 & 2.27 & radio selection & \markcite{Chambers:88} () -- HzRG record \\
%1987 & 3C~326.1     & 1.83 & radio selection & \markcite{McCarthy:87}McCarthy {et~al.} (1987) -- HzRG record \\ 
1984 & 3C~256       & 1.82 & radio selection & \markcite{Spinrad:84}Spinrad \& Djorgovski (1984) \\
1983 & 3C~324       & 1.206 & radio selection & \markcite{Spinrad:83}Spinrad \& Djorgovski (1983) \\
1982 & 3C~368       & 1.131 & radio selection & \markcite{Spinrad:82}Spinrad (1982) \\
%1982 & 3C~280      & 0.996 & radio selection & \markcite{Spinrad:82}Spinrad (1982) \\
%1981 & 3C~427.1     & 1.175 & radio selection & \markcite{Spinrad:81} () -- NOT \\
1979 & 3C~6.1       & 0.840 & radio selection & \markcite{Smith:79}Smith {et~al.} (1979) \\
%1979 & 3C~265      & 0.81 & radio selection & \markcite{Smith:79}Smith {et~al.} (1979) \\
%1977 & 3C~343.1    & 0.750 & radio selection & \markcite{Spinrad:77} () \\
1976 & 3C~318       & 0.752 & radio selection & \markcite{Spinrad:76}Spinrad \& Smith (1976) \\
%1975 July & 3C~123  & 0.637 & radio selection & \markcite{Spinrad:75b} () -- NOT \\
1975 & 3C~411   & 0.469 & radio selection & \markcite{Spinrad:75a}Spinrad {et~al.} (1975) \\
1960 & 3C~295       & 0.461 & radio selection & \markcite{Minkowski:60}Minkowski (1960) \\
1956 & Cl~0855+0321 & 0.20 & cluster selection & \markcite{Humason:56}Humason, Mayall, \& Sandage (1956) \\
\hline
\end{tabular}
\end{center}
\medskip

\emph{Notes.---} Status as of 1999 August. Tabulation restricted to
confirmed spectroscopic sources.  In particular, \markcite{Hu:98}Hu, Cowie, \& McMahon (1998) recently
reported a likely (serendipitously-discovered) candidate at $z = 5.63$
while \markcite{Chen:99}Chen {et~al.} (1999) report a candidate at $z = 6.68$ selected from
deep {\it HST}/STIS grism spectroscopy.  The authors deem both redshift
determinations tentative with the current data (see Stern \etal
1999b).  Note that \markcite{Petitjean:96}Petitjean {et~al.} (1996) refers to the spectroscopic
confirmation of the $z=4.7$ quasar companion initially identified by
\markcite{Djorgovski:95}Djorgovski (1995) and \markcite{Hu:96}Hu, McMahon, \& Egami (1996).  Many sources with potentially
higher photometric redshifts have been identified, but await
spectroscopic confirmation.

\label{gold}
\end{table}
\normalsize

Progress in this field has accelerated with the advent of new
facilities, notably the Keck telescopes.  In \S\ref{sectz5bcg} we
present a brief historical review of distant galaxy studies followed by
a discussion of protogalaxy searches at non-optical wavelengths in
\S\ref{sectz5notopt}.  In \S\ref{sectz5optir} we discuss several
optical/near-infrared selection techniques for the `normal' population
of distant galaxies.  The cosmological redshifting of the light from
these distant systems implies that our ground-based
optical/near-infrared window samples the rest-frame ultraviolet
spectrum;  in \S\ref{sectz5lowz} we therefore discuss the results of
recent space-based observations of the ultraviolet properties of the
youngest galaxies locally, as detailed studies of these relatively
bright systems can yield considerable insight into observations of the
most distant systems.  In \S\ref{sectz5bias} we discuss the biases in
the protogalaxy search techniques.  In \S\ref{sectz5results} we detail
some of the highlights of these studies.  Finally,
\S\ref{sectz5conclude} summarizes the discussion and suggests the
primary questions which may occupy workers in this field at the start
of the new millennium.

Throughout this paper, unless otherwise explicitly stated, we adopt an
Einstein-de~Sitter cosmology with a Hubble constant $H_0 = 50~ h_{50}~$
\kmsMpc and no cosmological constant, $\Lambda = 0$.  For this
cosmology, the age of the Universe at redshift $z$ is ${2 \over 3}~
H_0^{-1}~ (1+z)^{-3/2} = 13.2~ h_{50}^{-1}~ (1+z)^{-3/2}~$ Gyr.
Magnitudes are quoted in the Vega-based system unless otherwise
explicitly stated.

\section{Historical Background:  Brightest Cluster Galaxies}
\label{sectz5bcg}

%Identifying the rare, faint distant galaxy from the multitude of
%lower-luminosity field galaxies is a challenging endeavor.  Many
%techniques have been employed.  Until the advent of the $9 \pm 1$~m
%telescopes, brightest cluster members and radio selection were
%the primary sources of distant galaxies.  

Historically, identifying galaxy clusters from overdensities on deep
photographic plates was a proven method to find distant galaxies.
Because the central cD galaxy in a cluster is quite luminous relative
to the average galaxy ($M_V{\rm (cD)} \approx -24$ as opposed to $M_V^*
\approx -21.5$ for early-type galaxies in clusters), clusters are
excellent high-luminosity landmarks for studying the extragalactic
Universe and were used by astronomers from 1950 to 1980 to find
galaxies out to $z \approx 0.6$.  While no longer primary sites for
identifying distant galaxies directly because clusters require a
substantial fraction of the Hubble time to form and virialize, they are
once again becoming important for studying the process of structure
formation.  \markcite{Eke:96}Eke, Cole, \& Frenk (1996) show that the evolution of cluster
abundances is sensitive to basic cosmological parameters.  The temporal
evolution of the co-moving galaxy cluster number density is determined
by the rate of growth of large density perturbations.  This depends
mostly on the value of $\Omega$; in a low-density universe the cluster
population evolves slowly at low to intermediate redshift.  In a
critical ($\Omega = 1$) universe the density fluctuations continue to
grow to the present epoch --- thus, the cluster population is still
evolving and we should expect more clusters locally than at
intermediate redshift ($z \simeq 0.8$).  Preliminary results from deep
X-ray surveys do not indicate significant evolution in the galaxy
cluster population from $z \simeq 0.8$ to the local Universe,
suggestive of a low-$\Omega$ Universe \markcite{Rosati:00}(\eg Rosati 1999).

The gravitational lensing caused by massive galaxy clusters amplifies
the apparent brightness of background sources.  Several galaxies at $z
> 4$ have now been identified behind Abell clusters
\markcite{Frye:98}(\eg Frye \& Broadhurst 1998), with the most distant strongly-lensed source a
serendipitously-discovered system at $z =4.92$ \markcite{Franx:97}(Franx {et~al.} 1997).

\section{Searches at Non-Optical Wavelengths}
\label{sectz5notopt}

While optical and infrared techniques are usually needed to adequately
image distant galaxies and measure their redshifts, the candidates
may initially be located by observations in less-classical wavebands.
We discuss identifying galaxies at non-optical wavelengths in the
following subsections, beginning with radio selection which has
the richest history, and gradually moving to higher-energy photons.

%In particular we refer to sources found by radio surveys, and, at an
%increasing rate, those located by ground-based sub-mm emission and by
%space-borne high energy detectors in the X-ray and $\gamma$-ray
%domains.  Normal galaxies likely do not regularly radiate large
%fractions of their energies at these disparate wavelengths, but active
%galactic nuclei (AGN) and galaxies undergoing large episodes of star
%formation often do.  Gamma ray bursters (GRBs) are apparently
%associated with transient, highly energetic (beamed?) processes in
%distant galaxies.  These rare events can be used as useful beacons of
%extremely distant galaxies.  We note that sub-mm and far-infrared
%($\lambda > 60 \mu$m) sources may be associated with dusty, but
%otherwise normal galaxies at great distance, which are difficult to
%identify from optical selection alone as the observed optical at
%high-redshift samples rest-frame ultraviolet light and can be highly
%attenuated by dust absorption.

\subsection{High-Redshift Radio Galaxies}

Powerful radio galaxies have proven to be good targets as luminous,
presumably massive galaxies at large distance. They are spatially rare
and usually luminous over many decades of the electromagnetic spectrum.
As is clear from Table~\ref{gold}, until only recently, studies
of the highest-redshift galaxies were synonymous with studies of
high-redshift radio galaxies (HzRGs).

\begin{figure}[!ht]
\plotfiddle{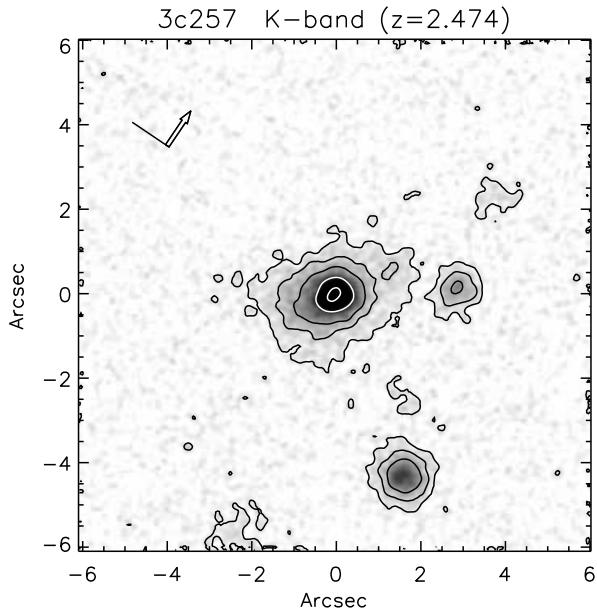}{3.2in}{0}{50}{50}{-130}{10}

\caption[Near-infrared image of 3C~257]{Near-infrared image of the HzRG
3C~257 at $z = 2.474$, obtained with the NIRC camera on Keck~I.  3C~257
is the highest-redshift radio galaxy identified from the 3CR survey.
Panel is 12\farcs0 square, centered on the HzRG and oriented such that
the inner radio axis is parallel to the abscissa.  The compass arrow
in the upper left indicates NE orientation, with north shown by the
heavier arrow.  The lowest contour corresponds to a surface brightness of
21.6 mag arcsec$^{-2}$.  Note that the morphology is well-described by
a deVaucoleurs $r^{1/4}$ profile, as expected for an early-type system.
Figure courtesy van Breugel \etal (1998).}

\label{3c257}
\end{figure}

After the pioneering radio-optical identifications by Baade, Minkowski,
Wyndham, Sandage, Ryle, Kristian, Longair, and Gunn, it became clear
that relatively nearby powerful radio galaxies ($P_{\rm 408 MHz} >
10^{28}\ {\rm W~Hz}^{-1}$) were usually associated with giant
elliptical (gE and cD) galaxies \markcite{Matthews:64}(Matthews, Morgan, \& Schmidt 1964).  Upon closer
inspection, a more precise description is that powerful radio sources
are identified with disturbed-looking ellipticals of high luminosity
(see Fig.~\ref{3c257}) and that optical absolute magnitude ($M_V$)
correlates with radio power ($P_{\rm 1.4 GHz}$) such that the brightest
early-type (giant elliptical) galaxies are much more likely to host a
radio source of significant power.  Could this trend be exploited to
find massive galaxies at high redshift?

% FIGURE 2

\begin{figure}[!t]
\plotfiddle{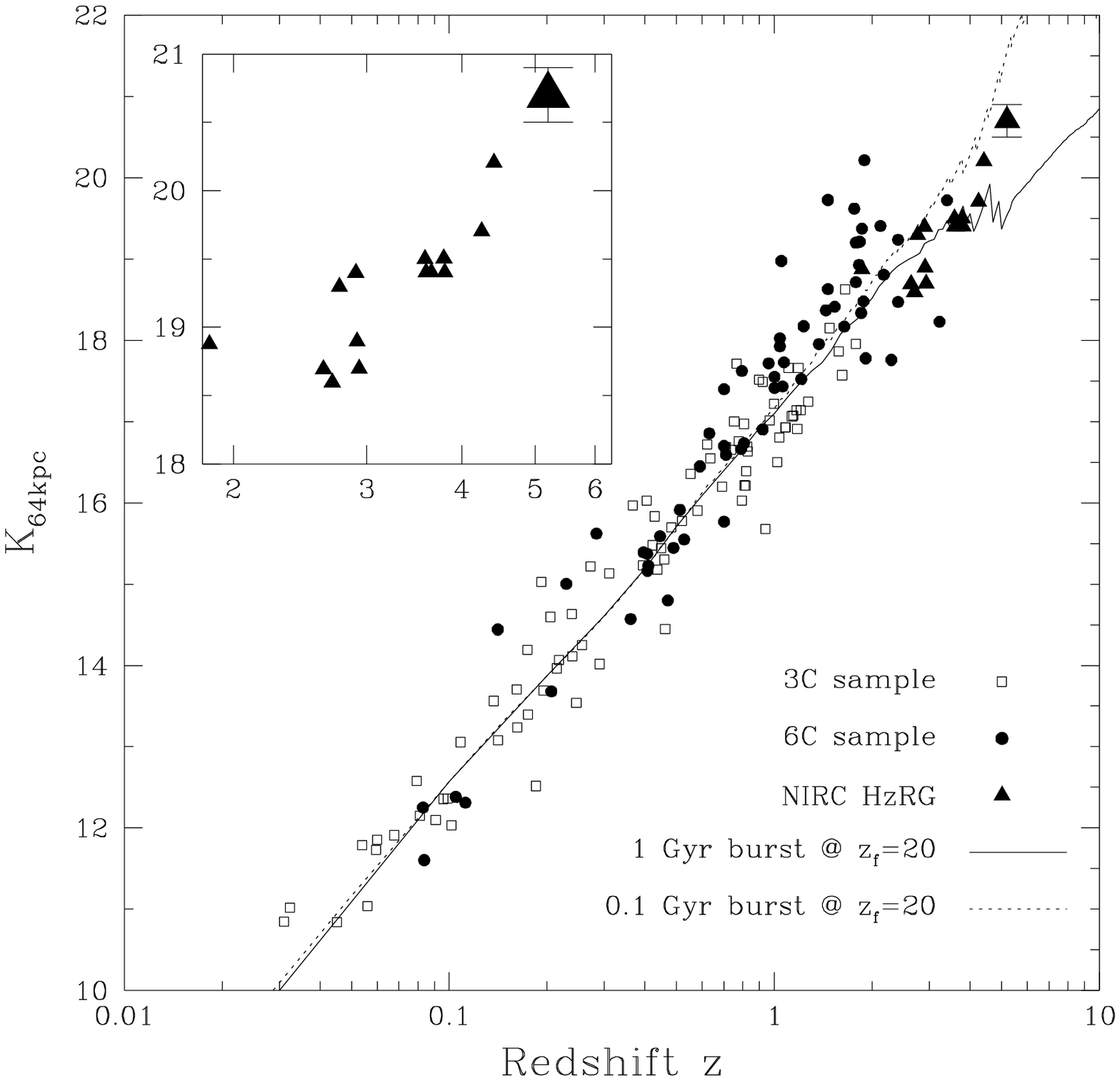}{3.0in}{0}{50}{50}{-170}{-100}

\caption[Hubble $K - z$ diagram for HzRGs]{Hubble $K - z$ diagram for
HzRGs.  Small filled triangles are Keck/NIRC measurements of HzRGs from
van Breugel \etal (1998), the large filled triangle with error bars
represents TN~J0924$-$2201 at $z = 5.19$ (van Breugel 1999), currently
the most distant HzRG known, and all other photometry is from Eales
\etal (1997).  Magnitudes are aperture-corrected to a 64~kpc metric
diameter using $H_0 = 65 \kmsMpc$ and $\Omega_0 = 0.3$.  The inset
highlights this empirical relation at the highest accessible redshifts
currently.  Two stellar evolutionary models from Bruzual \& Charlot
(1993) with formation redshifts $z_f = 20$ and normalized at $z < 0.1$
are plotted.  Figure courtesy van Breugel \etal (1999).}

\label{revkz}
\end{figure}

The first major effort concentrated on the hosts of radio sources from
the revised 3{\it rd} Cambridge catalog \markcite{Bennett:61}(3CR; Bennett 1961),
representing the brightest radio sources in the Northern hemisphere
($S_{\rm 178 MHz} > 9 {\rm Jy}$; $\delta > -5\deg$).  Follow-up
identifications by Longair, Gunn, Kristian, Sandage, Spinrad, and
collaborators \markcite{Djorgovski:88}(\cf Djorgovski {et~al.} 1988) has lead to nearly complete
spectroscopic redshift identification for the 328 radio galaxies in the
3CR catalogue.  Only one faint galaxy, the host of 3C~249, remains
without a definite spectroscopic redshift.  The highest redshift
belongs to 3C~257, a faint system ($K = 18.4$) at $z = 2.474$
\markcite{vanBreugel:98}(van Breugel {et~al.} 1998).  \markcite{Lilly:82, Lilly:84}Lilly \& Longair (1982, 1984) found a good
correlation between the 2.2$\mu$m ($K$-band) infrared magnitude and
galaxy redshift for powerful radio galaxies from the 3CR.  At the
redshifts of the 3CR, the $K$-band samples long-wavelength light,
thought to be less-heavily affected by young stars and AGN; it was
initially hoped that the $K$-band flux correlated tightly with mass and
that the HzRGs were tracking the evolution of early-type galaxies.
This discovery of a tight infrared Hubble, or $K - z$, diagram briefly
suggested that radio galaxies might be good standard candles even if
located at large redshift and thus become a viable tool for measuring
basic cosmological parameters, such as $q_0$.  Though this line of
research later proved unfruitful, HzRGs still generally obey the $K -
z$ relation out to the highest accessible redshifts currently probed
(see Fig.~\ref{revkz}), despite significant morphological evolution
\markcite{vanBreugel:98}(van Breugel {et~al.} 1998) and the dramatic $k$-correction effects ($K$
samples rest-frame $U$ at $z \sim 5$).

% FIGURE 3

To identify higher-redshift HzRGs and remove the redshift--radio power
degeneracy in studies of HzRGs, several contemporary efforts have
concentrated on lower flux density limit samples.  Recent mJy surveys
include the Molonglo Reference Catalog \markcite{McCarthy:96}(MRC;  McCarthy, Baum, \& Spinrad 1996),
the MIT-Green Bank survey \markcite{Stern:99a}(Stern {et~al.} 1999b), and Cambridge surveys
\markcite{Hales:93}(\eg 6C;  Hales, Baldwin, \& Warner 1993).  Additional efforts have concentrated on
$\mu$Jy samples such as the Leiden Berkeley Deep Survey
\markcite{Neuschaefer:95}(Neuschaefer \& Windhorst 1995) and sensitive radio maps of the Hubble Deep
Field \markcite{Richards:98}(Richards {et~al.} 1998).  These studies have shown that radio
luminosity weakly correlates with emission line luminosity
\markcite{Rawlings:89, Baum:89}(Rawlings {et~al.} 1989; Baum \& Heckman 1989), that weak radio galaxies do not generally
contain significant non-thermal light in their optical spectra
\markcite{Keel:91}(Keel \& Windhorst 1991), that the alignment effect (discussed below) becomes
less pronounced with lower radio flux density samples
\markcite{Rawlings:91, Dunlop:93, Eales:93}(\eg Rawlings \& Saunders 1991; Dunlop \& Peacock 1993; Eales \& Rawlings 1993), and that radio power
apparently correlates with ionization state in HzRGs
\markcite{Stern:99a}(Stern {et~al.} 1999b).

To identify the highest-redshift HzRGs, two techniques have recently
been exploited with considerable success, yielding the first detection
of an HzRG at $z > 5$:  TN~J0924$-$2201 at $z = 5.19$
\markcite{vanBreugel:99a}(van Breugel {et~al.} 1999).  First, empirically, HzRGs with ultra-steep
spectra (USS; for $S_\nu \propto \nu^\alpha$, $\alpha \simlt -1.3$) at
radio wavelengths tend to be at higher redshifts
\markcite{Chambers:90}(\eg Chambers {et~al.} 1990).  All HzRGs at $z > 3.8$ have been
identified by targeting USS samples.  A partial explanation for this
technique derives from the observation that the most powerful radio
galaxies locally have radio spectral energy distributions which steepen
with frequency \markcite{Carilli:99}(\eg Carilli \& Yun 1999).  The $k$-corrections
therefore imply that sources at increasing redshifts exhibit steeper
radio spectral indices at a given observed frequency.  The second
technique relies on selecting those USS radio sources whose hosts have
the faintest observed $K$ magnitudes and thus, according to the $K - z$
diagram (Fig.~\ref{revkz}), are likely associated with the most distant
systems.  \markcite{DeBreuck:99}{De~Breuck} {et~al.} (1999) have recently constructed a full-sky USS
sample ($\alpha_{\rm 365 MHz}^{\rm 1.4 GHz} < -1.3$).  TN~J0924$-$2201
at $z=5.19$, a 1\farcs2 radio double with $S_{\rm 4.85 GHz} = 8.6 \pm
0.5$ mJy, $\alpha_{\rm 365 MHz}^{\rm 1.4 GHz} = -1.63 \pm 0.08$, and $K
= 21.3 \pm 0.3$ (2\farcs1 diameter circular aperture), is among the
steepest-spectrum USS sources from the \markcite{DeBreuck:99}{De~Breuck} {et~al.} (1999) sample and
has among the faintest $K$-band magnitudes of the subsample with
infrared imaging \markcite{vanBreugel:99a}(van Breugel {et~al.} 1999).  Table~\ref{tabhzrg} lists
several physical parameters for all HzRGs at $z > 3.8$ in the
literature currently.  At the time of this writing, the authors are
aware of 22 HzRGs at $z > 3$, including five at $z > 4$ and one at $z >
5$.  We note, however, that $K$-band imaging is not the Rosetta Stone
of HzRG studies; spectroscopy of several HzRGs with faint $K$-band
identifications (\eg $K \simgt 20$) reveal lower redshift systems, thus
defying a simple one-to-one translation of $K$-band magnitude to
redshift \markcite{Eales:96}(\eg Eales \& Rawlings 1996).

% TABLE 2
\begin{table}[ht!]
\caption{Physical Parameters of High-Redshift Radio Galaxies at $z > 3.8$}
\footnotesize
\begin{center}
\begin{tabular}{lccccccl}
\hline\hline
Galaxy &
$z$ &
$\alpha_{\rm 365 MHz}^{\rm 1.4 GHz}$ &
$K$ &
$S_{\rm 1.4 GHz}$ &
$L_{\rm Ly\alpha}$ &
$W_{\rm Ly\alpha}^{\rm rest}$ &
Reference \\
& & & (mag.) &
(mJy) & &
(\AA) & \\
\hline
TN~J0924$-$2201 & 5.19 & $-1.63$ & 21.3 &   73 & 0.8 & $\simgt 115$ & \markcite{vanBreugel:99a}van Breugel {et~al.} (1999) \\
VLA~J123642+621331&4.42& $-0.94$ & 21.4 & 0.432& 0.1 & $\simgt 50$ & \markcite{Waddington:99}Waddington {et~al.} (1999) \\  
6C~0140+326     & 4.41 & $-1.15$ & 20.7 &   92 & 7.7  & 700 &  \markcite{Rawlings:96}Rawlings {et~al.} (1996) \\
8C~1435+63      & 4.25 & $-1.31$ & 20.1 &  498 & 2.3 & 670 & \markcite{Spinrad:95}Spinrad, Dey, \& Graham (1995) \\
TN~J1338$-$1942 & 4.11 & $-1.31$ & 20.3 &  123 & 14.4 & 200 & \markcite{DeBreuck:99}{De~Breuck} {et~al.} (1999) \\
4C~41.17        & 3.80 & $-1.25$ & 20.7 &  266 & 9.0 & 100 & \markcite{Dey:97}Dey {et~al.} (1997) \\
\hline
\end{tabular}
\end{center}
\medskip

\emph{Notes.---}  \lya\ luminosity, $L_{\rm Ly\alpha}$, is in units of
$10^{43} h_{50}^{-2} \ergs$.  $W_{\rm Ly\alpha}^{\rm rest}$ is the
rest-frame \lya\ equivalent width.  The discovery papaers for
8C~1435+63 and 4C~41.17 were \markcite{Lacy:94}Lacy {et~al.} (1994) and \markcite{Chambers:90}Chambers {et~al.} (1990)
respectively.

\label{tabhzrg}
\end{table}
\normalsize

The radio morphologies of HzRGs often have a double-lobed structure.
At low-redshift, optical images of radio galaxies often show signs of
recent mergers, as evidenced by multiple nuclei and subtle shells around
some symmetric early-type galaxies to dramatic tidal tails around others
\markcite{Rigler:92}(\eg Rigler {et~al.} 1992).  At $z \simgt 0.7$, the optical axis of the
host galaxy is typically aligned with the radio axis \markcite{Chambers:87,
McCarthy:87}(Chambers, Miley, \& van Breugel 1987; McCarthy {et~al.} 1987).  Optical and radio major axes usually differ by less than
20\deg.  This alignment is most noticeable in emission-line images:
nearly all of the emission line regions of HzRGs in the 3CR sample
are well-aligned with their radio axis \markcite{McCarthy:95}(McCarthy, Spinrad, \& van  Breugel 1995).  At $z >
0.7$ the rest-frame ultraviolet continua are also strongly aligned.
At observed near-infrared wavelengths, the alignment is less pronounced.
Lower-redshift radio galaxies often have regular $r^{1/4}$ profiles
in the $K$-band \markcite{Best:96}(\eg Best, Longair, \& R\"ottgering 1996), while at the high-redshift end,
\markcite{vanBreugel:98}van Breugel {et~al.} (1998) find peculiar $K$-band morphologies --- faint,
large-scale ($\sim 50$ kpc) emission often surrounding multiple-compact
components aligned with the radio axis.

One interpretation of the alignment effect is that it is dominated by a
blue-component, which has variously been attributed to emission from
young, hot stars, scattered light from an active galactic nucleus (AGN;
a buried quasar or a buried mini-quasar), or nebular continuum emission
from clouds excited by an obscured nucleus \markcite{Dickson:95}(Dickson {et~al.} 1995).
Spectropolarimetric studies have convincingly demonstrated that
extended, aligned ultraviolet (UV) emission in HzRGs at $z \sim 1 - 2$
is often strongly polarized, with the electric vector perpendicular to
the major axis of the UV emission \markcite{Jannuzi:95, Dey:96,
Cimatti:96, Cimatti:97}(\eg Jannuzi {et~al.} 1995; Dey {et~al.} 1996; Cimatti {et~al.} 1996, 1997).  These observations strongly indicate that
much of the observed UV light is AGN light which has been scattered
into our line of sight by dust and/or electrons in the ambient medium.
Other galaxies, notably 4C~41.17 at $z = 3.80$ \markcite{Dey:97}(Dey {et~al.} 1997),
exhibit no polarized continuum in the aligned emission, but do show
stellar absorption features (\eg \ion{S}{5}$\lambda$ 1502), indicating
that (jet-)induced star formation is important in some HzRGs.  The true
scenario is likely a combination of these processes.

The observed infrared, sampling the rest-frame visible or UV for large
redshifts, is thought to be dominated by starlight from the host galaxy;
deductions made from the $H$- or $K$-bands might then be compared to
other large galaxies of stars.  The robustness of this conclusion is still
unproven \markcite{McCarthy:99b}(\eg see  McCarthy 1999).  To settle the point we need to
observe stellar spectral features in the near-infrared, but spectroscopy
of faint ($K \simgt 19$) galaxies at $2 \mu$m is technically difficult,
even with the largest telescopes.  However, the uniformity of the $K-z$
diagram (Fig.~\ref{revkz}) suspiciously traces the rapid formation of
a massive galaxy of stars at high redshift, similar to
expectations for the formation and evolution of early-type galaxies.

We now ask a more difficult question:  can the study of distant radio
galaxies tell us about the youthful galaxy population in general?
Until recently, HzRGs were the only stellar systems known at $z > 3$.
However, modern photometric studies of deep imaging fields (discussed
in detail in \S\ref{sectz5optir}) have led to the discovery of a
population of normal, star-forming galaxies at comparably large
distances.  At $z \sim 3$ the HzRGs are rather larger and more luminous
than these field galaxies:  a typical powerful radio galaxy at $z \sim
3$ has $K \approx 19$, while the normal, star-forming, field galaxies
at $z \sim 3$ typically have $K \approx 22$, some 3 magnitudes
fainter.  At $z \sim 3$ the HzRGs are typically spatially extended by
$1 - 2$ arcseconds at $K$, while the young, field galaxies are
generally barely resolved in space-based (optical) images
\markcite{Giavalisco:96}(half-light radii of $0\farcs2 - 0\farcs3$; Giavalisco, Steidel, \&  Macchetto 1996).
Assuming the observed $K$-band light is dominated by stars in both
cases, it would take $\sim 15$ young field galaxies to match the
luminosity of a powerful radio galaxy at $z \sim 3$.  Could it be that
the HzRGs were the first objects to form in the early Universe?  If so,
what is the significance of the massive black hole that is thought to
reside at the HzRG nucleus and power the synchrotron emission at radio
wavelengths?  Is it primordial, or the result of galaxy evolution in
the early Universe?  Extending the frontiers of distant galaxy studies
of both the young, star-forming systems and the HzRGs to earlier
cosmic epochs will be a valuable enterprise.  Eventually, as we probe
to the epoch in which the (proto-)elliptical hosts of HzRGs are being
constructed by galaxy mergers, we might expect to see diminishing
systematic differences in the long-wavelength luminosities of HzRGs and
star-forming galaxies.

% FIGURE 4

\begin{figure}[!ht]
\plotfiddle{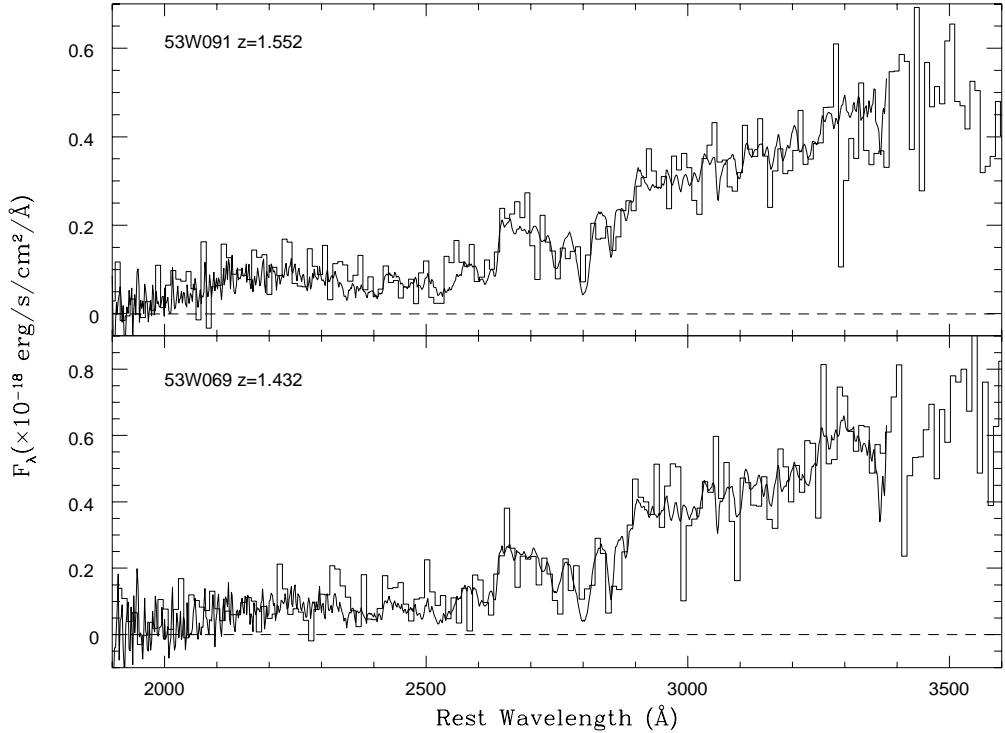}{3.4in}{-90}{52}{52}{-210}{295}

\caption[Keck spectra of the two oldest known high-redshift
galaxies]{Keck spectra of the two oldest known high-redshift galaxies.
The upper panel shows the spectrum of 53W091 (histogram) overlaid with
the mean spectrum of an F6V star from the Wu \etal (1991) {\it IUE}
Spectral Atlas.  The bottom panel shows the spectrum of 53W069
(histogram) compared with the mean spectrum of an F9V star. Since the
UV spectrum of coeval populations is dominated by light from
main-sequence turnoff stars, these fits imply ages of 3.5~Gyr for
53W091 and 4.5~Gyr for 53W069. These ages agree with those derived from
more-detailed population synthesis.  Figure courtesy \markcite{Dey:99c}Dey (1999).}

\label{oldreddead}
\end{figure}

Optical spectroscopy of a few (lower-power) HzRGs at $z \approx
1.5$ have shown spectra devoid of the narrow, high-equivalent width
emission lines which typically dominate HzRG spectra.  Detailed studies
of two galaxies \markcite{Dunlop:96,
Spinrad:97, Dey:99b}(LBDS~53W091 and LBDS~53W069 --- Dunlop {et~al.} 1996; Spinrad {et~al.} 1997; Dey {et~al.} 2000) show that their observed optical spectra are
well-represented by an evolved stellar population of age $\simgt 3.5$
Gyr (see Fig.~\ref{oldreddead}), with distinctive continuum decrements
evident at 2640 \AA\ and 2900 \AA\ whose amplitudes are similar to those
seen in late F-type stars.  Though the modeling of the rest-frame UV
spectra are uncertain currently --- different spectroevolutionary models
of a solar metallicity system with a delta function star formation history
yield significantly different inferred ages \markcite{Spinrad:97}(see Spinrad {et~al.} 1997) ---
the hypothesis that these systems have ages a substantial fraction of the
Hubble time at their observed redshifts seems unassailable.  These systems
are among our best examples of giant elliptical galaxies at $z > 1$.

The modest surface density of faint-/moderate-strength radio sources
reminds us of their rarity at any redshift:  the surface density of
radio sources with flux densities $S_{\rm 1.4 GHz} > 1$ mJy (\ie a
moderately weak radio flux) is $\simlt 0.03$ sources per square
arcminute \markcite{Becker:95}(Becker, White, \& Helfand 1995).  The co-moving density of HzRGs is
estimated to be $10^{-8} h_{50}^3$ galaxies Mpc$^{-3}$ at $z \sim 3$
\markcite{Dunlop:90}(\eg Dunlop \& Peacock 1990).  For reference, the co-moving density of
early-type galaxies locally is $\sim 2 \times 10^{-4} h_{50}^3$
Mpc$^{-3}$ \markcite{Cheng:99}(Cheng \& Krauss 1999).

Even if HzRGs prove not to be identified with an early stage in the
formation of all early-type galaxies, they remain convenient beacons
for large, overdense regions, populated by smaller normal galaxies and
are likely to be important for studies of the formation and evolution
of large scale structure.  Several galaxy clusters at $z \approx 1$
have been identified around distant radio galaxies
\markcite{Dickinson:95, Deltorn:97}(\eg Dickinson 1995; Deltorn {et~al.} 1997), with associated high-density
regions located even out to $z = 2.4$ \markcite{Pascarelle:96, Francis:96,
Pentericci:97, Carilli:98}(Pascarelle {et~al.} 1996; Francis {et~al.} 1996; Pentericci {et~al.} 1997; Carilli {et~al.} 1998).

For a review of HzRGs, see \markcite{McCarthy:93}McCarthy (1993).  We now consider
high-redshift galaxies selected at other non-optical/infrared
wavelengths.  Selecting targets at other wavelengths has the potential
to pick out the rare beast in the cosmos, which, from a Galilean world
view, is apt to be far away.

\subsection{Sub-mm Detections of Distant Galaxies}

Sub-mm identifications and observations of distant galaxies has
recently been energized by array detectors, most notably the
Sub-Millimeter Common User Bolometer Array \markcite{Holland:99}(SCUBA; Holland {et~al.} 1999)
on the 15-m James Clerk Maxwell Telescope on Mauna Kea.  SCUBA has
enabled the first deep, unbiased surveys to be made of the sub-mm sky
(generally implemented at 850$\mu$m).  Sub-mm observations are
particularly sensitive to dust emission, associated with re-processed
light from young, high-mass stars in galaxies undergoing massive bursts
of star formation.  The thermal dust spectrum peaks at a rest-frame
wavelength of around 100$\mu$m which has the interesting result of {\em
negative} sub-mm $k$-corrections:  the redshifted, modified black body
spectrum is sufficient to offset cosmological dimming for $\Omega_0 =
1$ and $1 \simlt z \simlt 10$!  Even for low values of $\Omega_0$, the
850$\mu$m flux density is only expected to decrease by a factor of a
few over this redshift range \markcite{Blain:93, Hughes:98}(Blain \& Longair 1993; Hughes {et~al.} 1998).  Dusty,
starburst galaxies can therefore be selected at $z \simgt 1$ in an
almost distance-independent manner.  Several relatively bright sources
have recently been identified \markcite{Smail:97, Barger:98,
Barger:99, Hughes:98}(\eg Smail, Ivison, \& Blain 1997; Barger {et~al.} 1998, 1999; Hughes {et~al.} 1998) and the flux densities may occasionally imply
immense star-formation rates in excess of 1000 M$_\odot$ yr$^{-1}$
\markcite{Dey:98}(\eg Dey {et~al.} 1998)!  A new question is what fraction of young
galaxies are dusty enough to re-emit strongly at $\lambda_{\rm rest}
\approx 100 \mu$m?  In particular, do the sub-mm galaxies constitute an
orthogonal population to the star-forming, UV-bright galaxies
considered in \S\ref{sectz5optir}?

%A crude limit on {\em heavy} dust shrouds covering most high-redshift
%galaxies is given by \markcite{Djorgovski:99} () with respect to the location
%of gamma-ray burst optical transients.

The main difficulty in the current interpretation of weak extragalactic
sub-mm sources is the relatively large (15\arcsec) beam size of SCUBA
at 850$\mu$m.  The issue is reminiscent of the early days of follow-up
work on radio sources:  sub-mm field sources have uncertain
identifications.  For example, the \markcite{Hughes:98}Hughes {et~al.} (1998) deep 850$\mu$m
observations of the Hubble Deep Field (HDF) detects five sources to a
limiting flux density of 2 mJy, corresponding to a surface density of
$\sim 0.9$ sub-mm sources per square arcminute to this limiting
brightness.  The sources are likely to be associated with individual
(starburst) galaxies over a large redshift range, with some
contribution of AGN likely.  The critical point is, of course, which
galaxies?  And what are their redshifts?  \markcite{Hughes:98}Hughes {et~al.} (1998) suggest
that one source (HDF850.4; $S_{\rm 850 \mu m} = 2.3 \pm 0.5$ mJy) is
associated with an extreme starburst galaxy (HDF~2-339.0; $I_{814}^{\rm
AB} = 23$) at $z \simeq 1$, whilst the other four sub-mm sources could
be identified with more distant galaxies, perhaps out to $z \approx
4$.  However, recent deep high-angular resolution radio observations of
the HDF by \markcite{Richards:99}Richards (1999) suggest alternate identifications of the
sub-mm sources:  translating the astrometric center of the sub-mm map
by 6\arcsec\ increases the number of radio identifications of sub-mm
sources from one out of five to four out of five.  If correct, this
would then suggest that the presently detectable sub-mm sources in the
HDF are restricted to $z \simlt 2$.

\markcite{Carilli:99}Carilli \& Yun (1999) show that the radio-to-sub-mm spectral index is a
viable redshift indicator for star-forming galaxies.  Using
semianalytic models and the well-studied local starburst galaxies M82
and Arp~220, they show that the 1.4~GHz to 850~$\mu$m (350~GHz)
spectral index $\alpha_{\rm 1.4~GHz}^{\rm 850 \mu m}$ increases with
redshift; galaxies with $\alpha_{\rm 1.4~GHz}^{\rm 850 \mu m} \simeq
0.5$ are likely to be at $z \simeq 1$.  Examining the \markcite{Hughes:98}Hughes {et~al.} (1998)
sub-mm identifications in the HDF, \markcite{Carilli:99}Carilli \& Yun (1999) conclude that
most of the sources are at $z \simeq 1.5$.  \markcite{Carilli:99}Carilli \& Yun (1999) also
show that the \markcite{Richards:99}Richards (1999) 6\arcsec\ offset yields low-redshift
identifications with inconsistently large radio-to-sub-mm spectral
indeces, implying that the \markcite{Hughes:98}Hughes {et~al.} (1998) astrometry is likely
accurate.

\markcite{Smail:97, Smail:98}Smail {et~al.} (1997, 1998) report on sub-mm selected sources from a
SCUBA cluster lens survey.  Deep sub-mm imaging of clusters offers two
advantages:  first, gravitational lensing magnifies any background sources
(a median amplification of $\sim 2.5$ is reported for this survey),
making spectroscopic follow-up easier.  Second, gravitational lensing
increases the mean separation of background sources, thus diminishing
source confusion.  \markcite{Barger:99}Barger {et~al.} (1999) reports on a spectroscopic follow-up
study of the possible optical counterparts and suggest that the
majority of the sub-mm sources reside at $z < 3$.  This implies that
the far-infrared and sub-mm background light recently measured from
the {\em FIRAS} and {\it DIRBE} experiments on the {\em COBE} satellite
\markcite{Schlegel:98}(\eg Schlegel, Finkbeiner, \& Davis 1998) are emitted by sources at $z < 3$.  Furthermore,
this work suggests that the peak activity in heavily obscured (dusty)
sources (both AGN and starbursts) lies at relatively modest redshift.

In the absence of high-angular resolution sub-mm capabilities, this
new field enjoys considerable uncertainties.  Further research is
clearly needed, and we can look forward to rapid progress in
understanding the redshift distribution, host galaxy properties, dust
properties, and evolution of the sub-mm population in the near future
as the next generation of mid-IR/sub-mm instruments (\eg SCUBA+,
BOLOCAM) become available.

\subsection{Luminous Infrared Galaxies}

The first all-sky, far-infrared survey, conducted in 1983 by the {\it
Infrared Astronomical Satellite} ({\it IRAS}), detected the existence of
a large population of galaxies which emit more energy in the infrared
($\sim 5 - 500 \mu$m) than at all other wavelengths combined.  In the
local Universe ($z \simlt 0.3$), these luminous infrared galaxies
(LIRGs) are the dominant population of sources with luminosities
above $10^{11} L_\odot$, being more numerous than quasars, Seyferts,
and optically-selected starbursts at comparable bolometric luminosity.
Morphological studies show that LIRGs are triggered by gas-rich galaxy
collisions and mergers.  The bulk of the infrared flux is powered by
dust heating from a massive starburst within giant molecular clouds.
At the highest luminosities, energy input from active galactic nuclei
becomes important, and LIRGs may represent an important stage in the
formation of quasars and powerful radio galaxies.  LIRGs may also be
an important phase in the formation of elliptical galaxies, globular
clusters, and in the metal enrichment of the intergalactic medium.
For a recent review of LIRGs, see \markcite{Sanders:96}Sanders \& Mirabel (1996).

In terms of studying distant galaxies, the {\it IRAS} sample of LIRGs
has not provided many sources.  The vast majority of {\it IRAS} LIRGs
are at relatively modest redshift ($z \simlt 0.4$).  Only two {\it
IRAS} sources have been identified at $z > 2$:  {\it IRAS} Faint
Source~10214+4724 at $z = 2.286$ \markcite{RowanRobinson:91}(Rowan-Robinson {et~al.} 1991) and the
Cloverleaf quasar at $z \sim 2.5$.  Both are gravitationally lensed
\markcite{Barvainis:94, Elston:94, Graham:95, Eisenhardt:96}(\eg Barvainis {et~al.} 1994; Elston {et~al.} 1994; Graham \& Liu 1995; Eisenhardt {et~al.} 1996), with
inferred amplification-corrected luminosities similar to the local
sample of LIRGs.  Studying the unlensed population of these
optically-obscured sources to higher redshift will be an important step
towards understanding the formation and evolution of massive galaxies,
but will require more sensitive infrared surveys and/or new search
strategies.  The {\it Space Infrared Telescope Facility} ({\it SIRTF})
is expected to identify LIRGs at cosmological distances.  Some sub-mm
sources are likely to be distant analogs of LIRGs.

\subsection{X-ray Emission Associated with Distant Galaxies}

Most normal galaxies are rather weak X-ray emitters, with spatially
extended X-ray emission in the range of $\sim 10^{38}$ \ergs\ to $\sim
10^{42}$\ergs\ \markcite{Fabbiano:89}(Fabbiano 1989).  This emission largely derives from
the hot phase of the interstellar medium (in early-type galaxies),
close accreting binaries, and the end products of stellar evolution.
Stellar coronal emission does not contribute significantly to the
total X-ray output of normal galaxies.  Extremely sensitive X-ray
surveys, such as the 56~ksec {\it R\"ontgensatellit} ({\it ROSAT}) Deep Survey in the Marano
field \markcite{Zamorani:99}(Zamorani {et~al.} 1999) and the 1.5~Msec {\it ROSAT} Deep Survey in
the Lockman Hole \markcite{Hasinger:98, Schmidt:98}(Hasinger {et~al.} 1998; Schmidt {et~al.} 1998), resolve most
of the hard X-ray ($0.5 - 2$~keV) background into discrete sources.
The majority of optically-identified X-ray sources are broad-lined AGN,
\ie quasars and Seyfert-II galaxies.  Recent claims of a new population of
X-ray emitting, narrow-emission line galaxies dominating faint X-ray
counts \markcite{Georgantopoulos:96}(Georgantopoulos {et~al.} 1996) are not supported by the deepest
surveys \markcite{Hasinger:98}(Hasinger {et~al.} 1998).

Thus, X-ray surveys are not expected to be significant tools for
identifying normal, distant galaxies, though they can identify distant
AGN.  The most-distant X-ray source known currently is the radio-loud
quasar GB~1428+4217 at $z=4.72$ identified due to its X-ray emission
\markcite{Fabian:97}(Fabian {et~al.} 1997).

X-rays are also emitted by bremsstrahlung radiation from the hot gas in
intracluster media.  Therefore, spatially-extended X-ray sources have
considerable value in locating distant, over-dense regions of the
Universe and are thus valuable probes for the formation and evolution
of large scale structure.  \markcite{Eke:96}Eke {et~al.} (1996) demonstrate that the
evolution of the abundance of rich clusters is strongly dependent upon
the cosmological density parameter $\Omega_0$.  Most X-ray clusters
identified to date lie at $z \simlt 0.5$.  \markcite{Stanford:97}Stanford {et~al.} (1997) and
\markcite{Rosati:99}Rosati {et~al.} (1999) report two spatially-proximate X-ray clusters at $z
\approx 1.27$.

The {\it Chandra X-Ray Observatory} (formerly known as {\em AXAF}) and
{\it X-Ray Multi-Mirror Mission} ({\em XMM}) will soon extend X-ray
detections to lower flux limits, providing a valuable aid in the
detection of very distant AGN and galaxy clusters.

\subsection{Gamma-Ray Bursts}

While the physical origin of gamma-ray bursts (GRBs) remains enigmatic,
it is now clear that they have an extragalactic origin and are
associated with an immense, possibly beamed, energy release.  This
makes them visible at large look-back times.  Several GRBs have been
credibly identified with X-ray, radio, and optical transients.  Most
spectacularly, GRB~990123 was identified with an optical transient
whose peak apparent brightness 47~s after the $\gamma$-ray release was
9$th$ magnitude in the optical, dimming by 5~mag within 500~s
\markcite{Akerlof:99}(Akerlof {et~al.} 1999).  As of this writing, nine GRBs have reliable
spectroscopic redshifts, with one event (GRB~971214) identified with a
galaxy at $z = 3.428$ \markcite{Kulkarni:98}(Kulkarni {et~al.} 1998).  Spectroscopy of the
GRB~990123 optical transient reveals an absorption system with $z_{\rm
abs} = 1.6004$ \markcite{Kelson:99}(Kelson {et~al.} 1999), setting a minimum distance for that
extremely bright source.  Assuming unlensed isotropic energy release, the
implied energies are immense (isotropic equivalent $E \simgt 10^{52 \pm
1}$ erg in the $\gamma$-rays alone).  Thus, at least some fraction of
GRBs are sufficiently powerful to be detected at great distances.  The
optical transients typically are not associated with the central nuclei
of galaxies, implying that GRBs are not related to AGN or the massive
black holes which are thought to reside in the centers of many galaxies
\markcite{Bloom:99}(Bloom {et~al.} 1999).  The two leading theories suggest that GRBs are
associated with the creation of a stellar-mass black hole, either
through coalescence of the remnants of a massive stellar binary
\markcite{Paczynski:86, Goodman:86}(\eg neutron star--neutron star or neutron star--black
hole;  Paczynski 1986; Goodman 1986) or through direct collapse of a
massive star \markcite{Woosley:93, Paczynski:98}(Woosley 1993; Paczynski 1998).  Both models predict
that GRBs should preferentially occur in star-forming galaxies.
Follow-up observations of the faint host galaxies of GRBs may become a
robust method of studying normal galaxies at the highest accessible
redshifts.  In particular, spectroscopy during the bright phase may
yield redshifts which would be untenable or impossible for the faint
hosts during their quiescent phase.

% FIGURE 5

\bigskip
\bigskip
\bigskip
\bigskip
\bigskip
\begin{figure}[!ht]
%\plotfiddle{fig5.eps}{4.5in}{0}{70}{70}{-160}{0}

\caption[Detail of images of the 0140+326 field]{{\bf \tt See
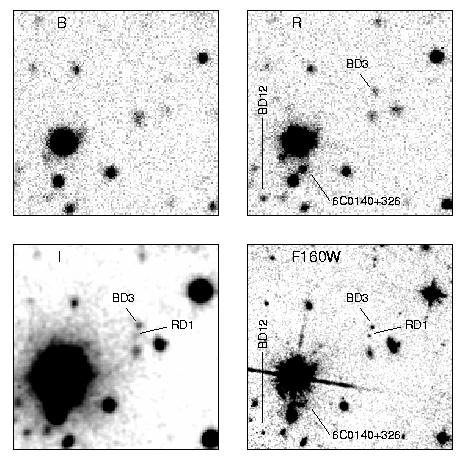.} Detail of the Keck/LRIS $BRI$ and {\it HST}/NICMOS F160W
($\sim H$) images of the 0140+326 field.  The images shown are
27\farcs5 on a side, and north is up and east is to the left.  Objects
at high redshift can be systematically and robustly selected on the
basis of redshifted Lyman line and continuum absorptions of hydrogen.
The galaxies targeted as $B$-band `dropouts' are labeled BD3 and BD12
and lie at $z = 4.02$ and $z = 3.89$ respectively.  The serendipitously
discovered $z = 5.34$ galaxy is labeled RD1 ($R$-band dropout; Dey
\etal 1998) in this figure.  6C~0140+326 (Rawlings \etal 1996) was the
most distant radio galaxy known at the time of the deep imaging and
lies at $z = 4.41$.  It is also a $B$-band dropout.  Hydrogen
absorption from the redshifted Lyman break and \lya\ forest causes
high-redshift objects to effectively disappear in the bluer passbands.}

\label{mosaic}
\end{figure}

\subsection{Summary}

High-redshift sources identified at non-optical/near-infrared
wavelengths therefore provide an important, albeit skewed, view of
galaxy formation, typically selecting unusual sources with powerful
active nuclei or systems undergoing massive bursts of star formation.
Recently several optical/near-infrared techniques have proved
successful at isolating the `normal' population of distant galaxies.
We discuss these methods below.  In \S\ref{sectz5bias} we discuss the
biases inherent in the various selection techniques, detailing how
populations identified at differing wavelengths compare.

\section{Optical/Near-Infrared Selection of Distant Galaxies}
\label{sectz5optir}

\subsection{Lyman-Break Galaxies}

In recent years our understanding of the earliest stages of galaxy
formation and evolution has been revolutionized.  With the commissioning
of the Keck telescopes atop Mauna Kea and the availability of deep
ground- and space-based imaging, the photometric technique outlined
in \markcite{Steidel:92}Steidel \& Hamilton (1992) has been used to routinely select high-redshift,
star-forming systems.  Dubbed the ``Lyman-break'' method, objects are
identified on the basis of the redshifted continuum spectral discontinuity
at $(1 + z) \times 912$ \AA\ and relatively flat (in $f_\nu$) continua
long-ward of the redshifted \lya\ (at $(1 + z) \times 1216$ \AA), as we
expect from hot O and B stars.  At the largest redshifts ($z \simgt 4$),
absorption due to the \lya\ and Ly$\beta$ forests plays a role of equal
or greater importance in attenuating the deep UV continua.  These hydrogen
absorptions from the Lyman limit and the Lyman forests cause high-redshift
objects to effectively disappear in the bluer passbands.  The technique
is therefore sometimes referred to as the `dropout' technique (see
Fig.~\ref{mosaic}).  $U$-band dropouts were initially used to
chart the Universe at $z \sim 3$ \markcite{Steidel:96a, Steidel:96b}(Steidel {et~al.} 1996a, 1996b).
This technique has now been pushed to higher redshifts: first to $z \simgt
4$ \markcite{Dickinson:98, Dey:98, Steidel:99}(Dickinson 1998; Dey {et~al.} 1998; Steidel {et~al.} 1999), and now finally above $z =
5$ \markcite{hdf4_473}( );][]{Weymann:98, Spinrad:98}.

% FIGURE 6

\begin{figure}[!ht]
\plotfiddle{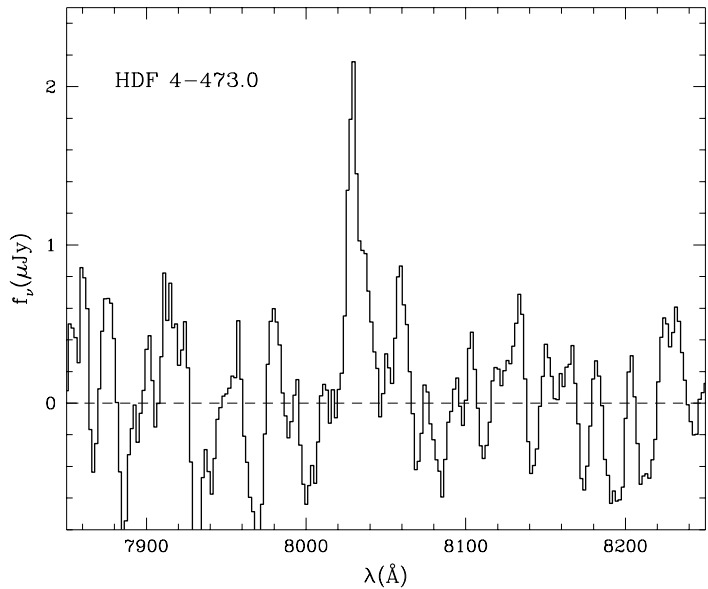}{2.8in}{0}{50}{50}{-150}{-100}

\caption[Spectrum of HDF~4-473.0]{Spectrum of HDF~4-473.0.  The total
exposure time is 4~h.  It was selected from the deep {\it HST} optical
and near-infrared images of the HDF (Williams \etal 1996; Thompson
\etal 1998) as a $V_{606}$-dropout ($V_{606} > 30$; $I_{814} = 27.1$
AB).  The spectrogram shows a single emission line at 8029 \AA.  The
asymmetric line profile and broad band colors are most consistent with
this being \lya\ at $z = 5.60$.  Figure courtesy Weymann \etal (1998).}

\label{hdf4_473}
\end{figure}

The hydrogen absorptions are ubiquitous:  they are present in the
spectra of O and B stars which will dominate the rest-frame UV continua
of young, star-forming galaxies with conventional mass functions.  They
are also present for any system with a substantial hydrogen column
present along our line of sight.  In particular, bound-free absorption
of hydrogen at the Lyman limit (at $(1 + z) \times 912$ \AA) will
severely truncate a background spectrum if the total neutral hydrogen
column at the relevant wavelength exceeds $n_H \approx 3 \times
10^{17}$ cm$^{-2}$.  Thus, Lyman absorption is expected even in systems
whose spectral energy distributions are not dominated by hot, young
stars; similar techniques have routinely been used to identify
high-redshift quasars from deep multi-band imaging such as the Palomar
Observatory Sky Surveys \markcite{Kennefick:95, Stern:99c}(\eg Kennefick, Djorgovski, \&  de~Calvalho 1995; Stern {et~al.} 1999c) and the
Sloan Digital Sky Survey \markcite{Fan:99}(Fan {et~al.} 1999).

\subsection{Photometric Redshifts}
\label{subsectz5photz}

Lyman-break and Lyman-forest methods are just two examples of
photometric redshift determinations, in which galaxy redshifts are
estimated using multiband photometric information.  The idea has a long
history, dating back to \markcite{Baum:62}Baum (1962) who used nine-band
photoelectric data to estimate galaxy cluster redshifts.
\markcite{Koo:85}Koo (1985), analyzing four-band photographic data, showed that
iso-redshift contours on color-color plots provide reliable photometric
redshift estimates.  Recently, photometric techniques have enjoyed a
revival, largely catalyzed by the deep photometry of the HDF and the
new generation of large-format CCD arrays.  The typical redshift
uncertainty for the photometric methods is expected to be in the range
of $\Delta z = 0.05 - 0.10$.  This accuracy is sufficient for many
scientific goals, such as luminosity function determinations,
luminosity density determinations, and projected correlation function
analyses.  The technique is also efficient for identifying
unusual sources, such as distant galaxies, quasars, and galaxy
clusters, which can then be targeted for detailed spectroscopic
study.

\begin{figure}[!ht]
\plotfiddle{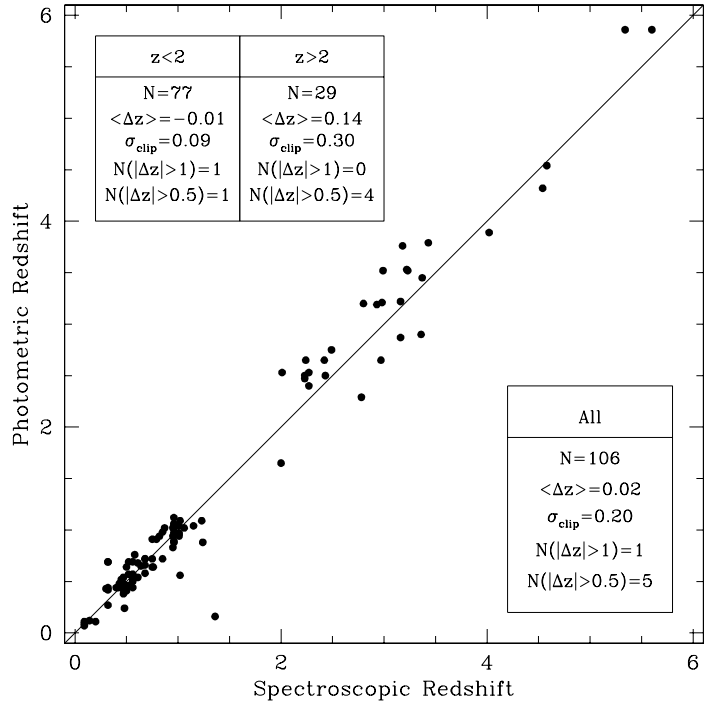}{3.2in}{0}{50}{50}{-150}{-90}

\caption[Spectroscopic vs. photometric redshifts in the HDF]{A
comparison of spectroscopic and photometric redshifts in the (Northern)
HDF.  Photometric redshifts are from the Stony Brook group, as detailed
in the text.  Spectroscopic redshifts are from the literature.  Insets
indicate various comparisons between the redshift determinations in two
redshift bins as well as the entire sample:  $N$ is the number of
galaxies, $\langle \Delta z \rangle$ is the average difference between
$z_{\rm spec}$ and $z_{\rm phot}$, $\sigma_{\rm clip}$ is the clipped
dispersion between the measurements, and $N(\vert \Delta z \vert >
0.5,1)$ measures the number of outlier photometric redshift
determinations.  Although this plot reports solely on the Stony Brook
photometric redshift determinations, dispersions of $\sigma \approx
0.2$ between $z_{\rm spec}$ and $z_{\rm phot}$ are typical of many of
the successful photometric redshift measurements in the HDF.  Figure
courtesy Fern\`andez-Soto.}

\label{specphotz}
\end{figure}

Several teams are actively pursuing photometric redshift determinations
and the HDF (recently augmented by the HDF-South) has been an excellent
laboratory for validating the technique \markcite{Hogg:98}(\cf Hogg {et~al.} 1998).
Table~\ref{photz} lists some of the primary working groups with brief
commentary on their technique.  Fig.~\ref{specphotz}, comparing
spectroscopic redshifts in the HDF with photometric redshift
determinations by the Stony Brook group, illustrates the robustness of
photometric redshift determinations.  A recent review of photometric
redshifts is presented in \markcite{Yee:98}Yee (1998).

% TABLE 3
\begin{table}[ht!]
\caption{Selected Groups Involved in Photometric Redshift Determinations}
\footnotesize
\begin{center}
\begin{tabular}{llc}
\hline\hline
Group &
Technique &
Ref. \\
\hline
Berkeley & Bayesian analysis with empirical spectral templates & 1 \\
Imperial College & synthetic spectral templates & 2 \\
Johns Hopkins & empirical fit to 4-dimensional flux space & 3 \\
Princeton & empirical fit to 3-dimensional color space &  4 \\
Stony Brook & hybrid spectral templates &  5 \\
Toronto & hybrid spectral templates & 6 \\
Victoria & synthetic and empirical spectral templates & 7 \\
\hline
\end{tabular}
\end{center}
\medskip

\emph{Notes.---}
References:
(1) \markcite{Benitez:99}Ben\'{\i}tez (1999);
(2) \markcite{Mobasher:96}Mobasher {et~al.} (1996); 
(3) \markcite{Connolly:95, Connolly:97, Brunner:97}Connolly {et~al.} (1995, 1997); Brunner {et~al.} (1997);
(4) \markcite{Wang:98}Wang, Bahcall, \& Turner (1998); 
(5) \markcite{Lanzetta:96, FernandezSoto:99}Lanzetta, Yahil, \&  Fern\`andez-Soto (1996); Fern\`andez-Soto, Lanzetta, \&  Yahil (1999); 
(6) \markcite{Sawicki:97}Sawicki, Lin, \& Yee (1997); 
(7) \markcite{Gwyn:96}Gwyn \& Hartwick (1996).

\label{photz}
\end{table}
\normalsize

% FIGURE 7

Contemporary photometric redshift determinations can be divided into two
major strategies.  The first approach \markcite{Koo:85, Lanzetta:96,
Gwyn:96, Mobasher:96, Sawicki:97, Benitez:99, FernandezSoto:99}(\eg Koo 1985; Lanzetta {et~al.} 1996; Gwyn \& Hartwick 1996; Mobasher {et~al.} 1996; Sawicki {et~al.} 1997; Ben\'{\i}tez 1999; Fern\`andez-Soto {et~al.} 1999) involves
fitting observed galaxy colors with redshifted spectral templates.  These
templates may be empirical, synthetic, or hybrid, and may be modified by
dust absorption and/or the redshift-dependent opacity of intergalactic
hydrogen.  Statistical treatments are used to determine a redshift
distribution function, generally quoted as a single number corresponding
to the peak of the distribution.  However, determinations of, for
example, galaxy luminosity functions, should retain the photometric
redshift distribution function to reliably indicate the uncertainties of
the analysis.  The other approach \markcite{Connolly:95, Brunner:97,
Connolly:97, Wang:98}(\eg Connolly {et~al.} 1995; Brunner {et~al.} 1997; Connolly {et~al.} 1997; Wang {et~al.} 1998) is purely empirical:  with a sufficiently large
training set, an empirical relation between redshift and observed
magnitudes $m_0$ and colors $C$ can be determined, $z = z (m_0, C)$.

We detail the method used by the Stony Brook group \markcite{Lanzetta:96,
FernandezSoto:99}(Lanzetta {et~al.} 1996; Fern\`andez-Soto {et~al.} 1999) to give some flavor of photometric redshift
determinations.  They begin with the four empirical galaxy spectral energy
distributions of \markcite{Coleman:80}Coleman, Wu, \& Weedman (1980), corresponding to four observed
galaxy types (E/S0, Sbc, Scd, and Irr) and covering the wavelength
range $1400 - 10,000$ \AA.  The templates are extrapolated in the UV
to 912 \AA\ using the empirical spectra of \markcite{Kinney:93}Kinney {et~al.} (1993), and are
extrapolated in the infrared to 25 $\mu$m using the spectral evolutionary
models of \markcite{Bruzual:93}Bruzual \& Charlot (1993).  The hybrid templates are redshifted and
the redshift-dependent ultraviolet hydrogen absorptions are removed
using a model of the optical depth of hydrogen.  At modest redshift,
this transmission function is empirically derived from observations
of quasars \markcite{Madau:95}(\eg Madau 1995).  The shifted, absorbed spectra are
then convolved with the transmission curves of the relevant filters and
a redshift likelihood function is calculated relating the model colors
to measured fluxes.  Comparisons between photometric and spectroscopic
redshift determinations in the HDF show that the former is typically
robust to $\Delta z \approx 0.34$ for objects with $I_{814} \sim 25.5$
and $z > 3$.  Residuals are typically $\Delta z_{\rm rms} / (1 + z)
\approx 0.1$ at all redshifts (see Fig.~\ref{specphotz}).

%The technique of \markcite{Giallongo:98} () is similar, but with two important
%differences.  First, rather than using a library of empirical template
%spectra, they construct synthetic templates utilizing the models of
%\markcite{Bruzual:93}Bruzual \& Charlot (1993).   Second, their aim is to determine photometric
%redshifts in the Giallongo 1202 field.  Therefore, they calibrate their
%redshift determinates using the broad-band colors and spectroscopic
%redshifts of 55 galaxies in the HDF.  The results of these photometric
%redshift predictions are, in general, quite satisfactory over the intended
%ranges of redshift applications.

Other template photometric redshift estimates generally vary in only two
considerations:  (1) the input spectral templates and (2) the method of
determining the best-fit $z_{\rm phot}$.  Input spectral templates may
be purely empirical, purely synthetic, or hybrid as in the Stony Brook
analysis (see Table~\ref{photz}).  Input libraries may also vary in
the number of templates.  Many groups use the four empirical spectral
energy distributions of \markcite{Coleman:80}Coleman {et~al.} (1980).  The addition of one or
more star-forming galaxy templates, as assembled by \markcite{Calzetti:94}Calzetti, Kinney, \&  Storchi-Bergmann (1994),
has been noted to improve the accuracy of the photometric redshift
determination by some groups.  Determination of the best-fit $z_{\rm
phot}$ relies on a statistical analysis.  For example, the Stony Brook
group uses a maximum likelihood analysis, while \markcite{Benitez:99}Ben\'{\i}tez (1999)
applies Bayesian marginalization and prior probabilities to the problem.

The training set method has the advantage of being purely empirical,
and therefore not relying on a choice of input templates.  However,
the weakness is that the method requires and is dependent upon a large
and accurate input training set.  Redshift ranges with sparse numbers
of confirmed redshifts, such as $1 \simlt z \simlt 2.5$ and $z \simgt
4$, lead to poorly determined photometric redshift determinations.
The various practitioners of this technique vary in the degree polynomial
used to fit the multivariate function $z = z(m_0, C)$ and whether or
not they implement observed flux as a variable in the optimal fit.

\begin{figure}[!ht]
\plotfiddle{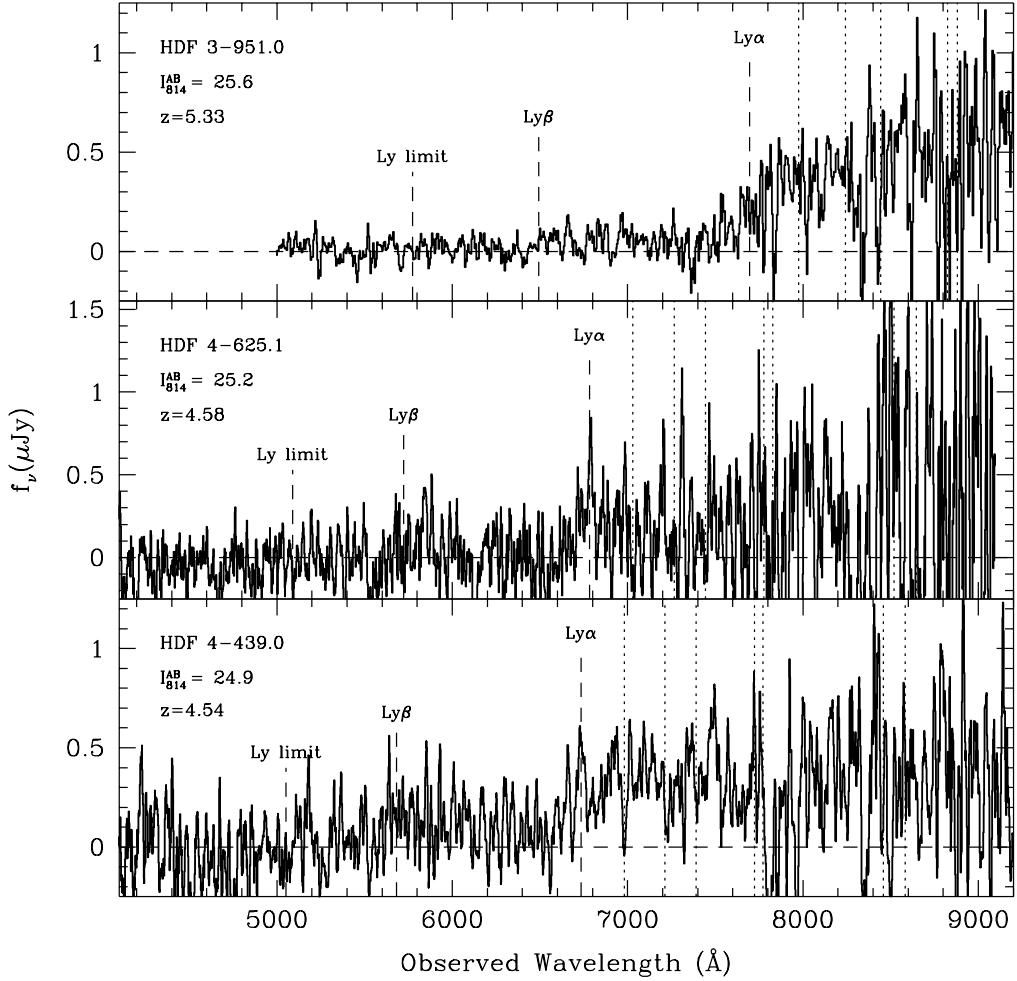}{4.9in}{0}{70}{70}{-222}{-100}

\caption[Keck/LRIS spectra of high-redshift galaxies in the
HDF]{Keck/LRIS spectra of HDF~3-951.0, HDF~4-625.0, and HDF~4-439.0,
three photometrically-selected high-redshift galaxies in the HDF.  The
spectrum of HDF~3-951.0 represents 6~hours of integration with the
150/7500 lines mm$^{-1}$ grating, significantly more sensitive than the
discovery spectrogram presented in Spinrad \etal (1998).  The
slightly-revised redshift of $z = 5.33$ is based on the Ly$\beta$
discontinuity, which is located at a wavelength less affected by
telluric emission than the \lya\ discontinuity.  The spectra
of HDF~4-625.0 and HDF~4-439.0 represent sums of data from multiple
observing seasons.  Vertical dotted lines indicate the location of
absorption lines typically observed in Lyman-break galaxies at $z
\simeq 3$.}

\label{hdf2spec}
\end{figure}

In terms of the current review on search techniques for protogalaxies,
we are most interested in the robustness of these photometric procedures
in identifying the high-redshift ($z \simgt 4$) tail in the field galaxy
redshift distribution.  \markcite{FernandezSoto:99}Fern\`andez-Soto {et~al.} (1999) find 5 galaxies brighter
than $I_{814}=26.5$ in the HDF with $4.5 < z < 5.5$, corresponding to
a surface density of 1.0 galaxies per square arcminute per unit $z$ at
$z = 5$.  Spectroscopic confirmation of the high-redshift photometric
candidates is necessary to validate the technique before photometric
redshift measurements of, for example, the star-formation history of the
early Universe, are well established.  Over the past three years, our
Berkeley-based group has been using slitmasks with the Low Resolution
Imaging Spectrometer \markcite{Oke:95}(LRIS; Oke {et~al.} 1995) on the Keck~II telescope
to measure faint galaxy spectra in the HDF.  We choose $z \simgt 4$
candidates in collaboration with the Stony Brook-based photometric
redshift group.  With the current instrumentation, spectroscopic redshift
measurements are viable perhaps to $I_{814} \approx 27.5$ for emission
line sources and to $I_{814} \approx 26$ for objects without strong
emission lines.  But even at $I_{814} \approx 25.5$, the effort is
fairly ``heroic'': observations of a single faint galaxy can easily
extend over multiple observing seasons in order to measure a reliable
spectroscopic redshift.  Fig.~\ref{hdf2spec} presents Keck/LRIS spectra
of three high-redshift sources in the HDF obtained by our Berkeley-based
group.  In Table~\ref{zspecphot} we list a comparison of spectroscopic
and photometric redshifts for galaxies in the HDF at $z > 4$.

% FIGURE 8

% TABLE 4
\begin{table}[ht!]
\caption{Spectroscopic vs. Photometric Redshifts in the HDF at $z > 4$}
\footnotesize
\begin{center}
\begin{tabular}{lccccl}
\hline\hline
Galaxy &
$z_{\rm spec}$ &
$z_{\rm phot}$ &
$W_{\rm Ly\alpha}^{\rm obs}$ &
$I_{814}$ (AB) &
Reference \\
& & & (\AA) & (mag.) & \\
\hline
HDF~3$-$512.0 & 4.02 & 3.56 & $\simgt 200$ & 25.5 & \markcite{Dickinson:98}Dickinson (1998) \\
%HDF~4$-$200.0 & 4.2: & 4.76 & & 27.0 & unpublished data \\
HDF~4$-$439.0 & 4.54 & 4.32 & $\sim 20$ & 24.9 & \markcite{Stern:00}Stern {et~al.} (2000) \\
HDF~4$-$625.0 & 4.58 & 4.52 & $\sim 100$ & 25.2 & \markcite{Stern:00}Stern {et~al.} (2000) \\
HDF~3$-$951.0 & 5.33 & 5.72 & \nodata & 25.6 & \markcite{Spinrad:98}Spinrad {et~al.} (1998) \\
HDF~4$-$473.0 & 5.60 & 5.64 & $\sim 300$ & 27.1 & \markcite{Weymann:98}Weymann {et~al.} (1998) \\ 
\hline
\end{tabular}
\end{center}
\medskip

\emph{Notes.---}  Photometric redshift $z_{\rm phot}$ from \markcite{FernandezSoto:99}Fern\`andez-Soto {et~al.} (1999).  $I_{814}$ magnitude from \markcite{Williams:96}Williams {et~al.} (1996).

\label{zspecphot}
\end{table}
\normalsize

The reliability of the photometric redshifts in the $4 \simlt z \simlt
5$ regime seems excellent.  Even at lower redshifts, only $\approx$
5\%\ of the photometric redshifts are noticeably discordant when
compared to spectroscopic redshifts.  However, looking ahead, as we
push the frontier past $z = 6$, the hydrogen absorptions are predicted
to almost completely obliterate the observed optical flux.  Deep
near-infrared photometry will be necessary, observations which are best
done from space, above the atmospheric water absorptions.  The NICMOS
camera on the {\it Hubble Space Telescope} ({\it HST}) has left a
valuable legacy of data and the Next Generation Space Telescope ({\it
NGST}) is being planned to work long-ward of 1$\mu$m.

%\markcite{FernandezSoto:99}Fern\`andez-Soto {et~al.} (1999) find 5 galaxies brighter than $I_{814}=26.5$
%in the HDF with $4.5 < z < 5.5$, corresponding to a surface density of
%1.0 galaxies per square arcminute per unit $z$ at $z = 5$.  Substantial
%dangers arise in deriving quantities such as the co-moving space density
%of normal, star-forming galaxies at $z = 5$ from one or two relatively
%small (a few square arcminutes) regions of sky.  However, if we take
%the \markcite{Steidel:99}Steidel {et~al.} (1999) empirical Lyman-break galaxy luminosity function
%derived from $\sim 600$ $U$-drop galaxies at $z \sim 3$ and assume
%a constant ($z$-independent) co-moving space density of star-forming
%galaxies at higher redshifts, as suggested by \markcite{Steidel:99}Steidel {et~al.} (1999), we
%can predict the surface density of galaxies at higher redshifts as a
%function of apparent magnitude.  For a critical ($q_0 = 0.5$) cosmology,
%we anticipate a surface density of $\approx 1$ galaxy per square arcminute
%at $z > 4.5$ and $I_{814} \leq  26$ (AB).  If we extend our limit to
%$I_{814} \leq  27$ (AB), the predicted surface density increases by a
%factor of $\approx 4$.  For $z > 5.5$, the predicted surface densities
%are 0.1 (0.5) galaxies per square arcminute for $I_{814} \leq 26 (27)$
%(AB).  This issue is further developed in \S 7.

\subsection{Emission Lines Searches for Distant Galaxies}

The stellar population synthesis models of \markcite{Charlot:93}Charlot \& Fall (1993) predict
rest-frame \lya\ equivalent widths of $50 - 200$ \AA\ for young, dust-free
galaxies.  An unattenuated \lya\ equivalent width of 200 \AA\ corresponds
to $\approx 5$\%\ of the bolometric luminosity of the protogalaxy being
emitted as \lya\ photons.  For a constant star formation history, the
\lya\ luminosity and equivalent width are only somewhat dependent on the
star formation rate and are greatest at times less than 10~Myr after the
onset of the burst.  Although this line emission is quite vulnerable to
extinction from dust, little dust might be expected in the first phase of
star formation.  Indeed, \markcite{Thommes:98}Thommes {et~al.} (1998) suggest requiring strong \lya\
emission as a necessary criterion for primeval galaxies.  Alternatively,
dust production can proceed remarkably quickly in supernovae remnants,
implying that most ``primordial'' galaxies may contain substantial dust by
the time we observe them.  We note that recent ultraviolet studies
of low-redshift \lya\ suggest that the kinematics of neutral gas may
be the more important attenuator of line emission \markcite{sectz5lowz}( )]{Kunth:98a}.

Several programs are currently underway to search for high-redshift
primeval galaxies through deep narrow-band imaging.  The previous
generation of surveys failed to confirm any field \lya-emitting
protogalaxy candidates \markcite{Thompson:95, Pritchet:94}(Thompson \& Djorgovski 1995; Pritchet 1994).  One of the
new programs is the Calar Alto Deep Image Survey
\markcite{Thommes:98}(CADIS; Thommes {et~al.} 1998) which uses a Fabry-P\'erot interferometer
to search for emission lines in three windows of low night sky emission
corresponding to \lya\ at $z =$ 4.75, 5.75, and 6.55.  The survey will
eventually sample 0.3 deg$^2$ of sky to a flux density limit of $S_{\rm
lim}(5 \sigma) \approx 5 \times 10^{-17} \ergcm2s$.  The technique
employs an arsenal of narrow-band `veto' filters to distinguish
foreground emission-line galaxies from distant \lya\ emitters.
Medium-band filters are also used to discriminate foreground objects
from \lya-emitting protogalaxy candidates at high redshift on the basis
of spectral energy distributions.  Six early candidates were reported
in \markcite{Thommes:98}Thommes {et~al.} (1998).  However, follow-up spectroscopy with the
Keck~II telescope has not confirmed a \lya\ interpretation for any of
these sources; \markcite{Thommes:99}Thommes (1999) present a revised candidate surface
density of 0.2 \lya\ emitters per square arcminute per unit-$z$ at $z =
5.75$ to the relatively-bright CADIS survey flux-density limit.

Another program to identify strong \lya-emitting star-forming galaxies
recently has been started at the Keck~II telescope \markcite{Cowie:98,
Hu:98}(Cowie \& Hu 1998; Hu {et~al.} 1998).  Using a combination of narrow-band interference filter imaging
and broad-band imaging, they discriminate high-redshift \lya\ emitters
on the basis of high equivalent width ($W_\lambda^{\rm obs} > 77$ \AA)
and broad-band color.  The survey probes to a flux density limit of
$S_{\rm lim}(5 \sigma) \approx 1.5 \times 10^{-17} \ergcm2s$.
Preliminary results, covering 46 arcmin$^2$ with a 5390/77 \AA\ filter,
suggest a surface density of $\approx 3$ \lya\ emitters per square
arcminute per unit-$z$ at $z \sim 3.4$ \markcite{Cowie:98}(Cowie \& Hu 1998).  The Hawaii
group has narrow-band filters tuned to gaps in the telluric OH
emission, corresponding to redshifts $z = 3.4, 4.5, 5.8$ and $6.5$.

Serendipitous searches on deep slit spectra, as discussed in the
following subsection (\S\ref{subsectz5ser}), are also sensitive to
line emission.  A technique combining narrow-band filters and
spectroscopy is discussed in \S\ref{subsectz5nbspec}.

Of course, identification of a strong emission line alone does not
necessarily imply the detection of high-redshift \lya.  Arguments based
upon the equivalent width of the line, lack of other emission lines,
the line profile, and associated continuum decrements are typically
used to determine the line identity \markcite{Stern:99b}(\eg Stern {et~al.} 1999a).  However,
selecting objects on the basis of strong emission samples a different
galaxy population from the traditional magnitude-limited surveys:
emission-line surveys are much more sensitive to active galaxies and
objects undergoing massive bursts of star formation.  Distinguishing
the redshift and source of line emission is challenging; comparison to
field surveys selected on the basis of continuum magnitude is perhaps
inappropriate.

For example, \markcite{Stern:99b}Stern {et~al.} (1999a) recently reported the serendipitous
detection of an emission line at 9185 \AA\ with an observed frame
equivalent width $> 1225$ \AA\ (95\%\ confidence limit).  The spectral
atlas of nearby galaxies by \markcite{Kennicutt:92}Kennicutt (1992) shows that the rest-frame
equivalent width of the H$\alpha$ + [\ion{N}{2}] complex rarely exceeds
200 \AA, that of [\ion{O}{3}]$\lambda$ 5007 \AA\ rarely exceeds
100 \AA, H$\beta$ rarely exceeds 30 \AA, and [\ion{O}{2}]$\lambda$
3727 \AA\ (hereinafter [\ion{O}{2}]) rarely exceeds 100 \AA.  Field
surveys of moderate-redshift, star-forming galaxies substantiate that
[\ion{O}{2}] rarely has a rest-frame equivalent width exceeding 100 \AA\
\markcite{Songaila:94, Guzman:97, Hammer:97, Hogg:98}(\eg Songaila {et~al.} 1994; Guzm\`an {et~al.} 1997; Hammer {et~al.} 1997; Hogg {et~al.} 1998).  Preliminary
analysis would strongly suggest that this source was a \lya-emitter at
$z = 6.56$, for which the implied rest-frame \lya\ equivalent width
would be consistent with confirmed sources at $z > 5$ \markcite{Dey:98,
Weymann:98}(Dey {et~al.} 1998; Weymann {et~al.} 1998).  However, a long exposure Keck/LRIS spectrogram revealed
a source 2\farcs7 away with strong [\ion{O}{2}] emission offset by only
7 \AA\ spectrally, persuasively arguing that this is an unusual, likely
active, [\ion{O}{2}]-emitting system at $z = 1.46$.

Infrared programs, similar to these optical programs, have begun to
yield some candidates, most likely at intermediate redshift thus far.
\markcite{Malkan:95}Malkan, Teplitz, \& McLean (1995) and \markcite{Teplitz:98}Teplitz, Malkan, \& McLean (1998), using the NIRC camera on
the Keck~I telescope, detect H$\alpha$ (but possibly [\ion{O}{2}]
or [\ion{O}{3}]) emission using the 2.16$\mu$m narrow-band CO filter.
These initial searches were targeted; \ie fields were selected to search
for spatially-correlated emission line galaxies around known quasars or
damped quasar absorption systems.

\markcite{McCarthy:99c}McCarthy {et~al.} (1999) report on blank-sky grism searches obtained during
parallel time with the NICMOS camera on {\it HST}.  The observations
sample $\approx 1.1 - 1.9 \mu$m and are unaffected by atmospheric water
absorption bands.  They find a surface density of single emission line
galaxies (most likely H$\alpha$) of $\approx 0.5$ galaxies per square
arcminute to a limiting flux density of $2 \times 10^{-17}$ \ergcm2s.
No variation with wavelength (redshift) is statistically significant in
their data set, which corresponds to H$\alpha$ emission over the
redshift range $0.7 \simlt z \simlt 1.9$.  Recombination theory coupled
with an assumed stellar luminosity function can be used to relate
H$\alpha$ luminosity to the star-formation rate $\dot{M}$
\markcite{Kennicutt:83, Madau:98}(\eg Kennicutt 1983; Madau, Pozzetti, \& Dickinson 1998).  \markcite{Yan:99}Yan {et~al.} (1999) find an average
star-formation rate of $\dot{M} = 21~h_{50}^{-2}~M_\odot$ yr$^{-1}$.

% FIGURE 9

\bigskip
\bigskip
\bigskip
\bigskip
\bigskip
\begin{figure}[!ht]
%\plotfiddle{fig9.eps}{2.6in}{0}{100}{100}{-200}{-10}

\caption[Two-dimensional spectrogram of serendipitous line
emitters]{{\bf \tt See 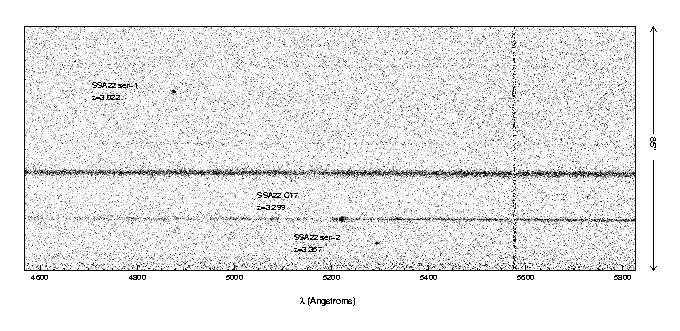.}Two-dimensional spectrogram of
serendipitously-discovered strong line emitters in the SSA22 field.
The source SSA22-C17 is a Lyman-break galaxy at $z = 3.299$ discovered
by Steidel and collaborators.  During deep, moderate-dispersion
Keck/LRIS follow-up spectroscopy of that source, two strong
line-emitters were serendipitously discovered on the same multislit
slitlet.  The high-equivalent widths, lack of secondary emission
features, and narrow velocity width of the lines argue that these are
high-redshift \lya-emitters.  Figure courtesy Manning \etal (2000).}

\label{serendips}
\end{figure}

\subsection{Serendipitous Longslit Searches }
\label{subsectz5ser}

Deep spectroscopy alone is also an efficient means to detect \lya\
emission from very high-redshift systems (see Fig.~\ref{serendips}).
Long spectroscopic integrations are sensitive to line-emitting sources
which serendipitously fall within the slit, potentially out to $z
\approx 6.5$.  This limit is set by the plummetting response of CCDs in
the near-infrared and to some degree by the strong OH sky emissions at
$\lambda > 9300$ \AA; infrared spectrographs can potentially extend
these surveys to still higher redshifts.  The first confirmed object at $z
> 5$ was the result of a serendipitous detection
\markcite{rd1}( );][]{Dey:98}.  \markcite{Hu:98}Hu {et~al.} (1998) also report the
serendipitous detection of an isolated emission line source which they
interpret as \lya\ at $z > 5$.  Serendipitous longslit searches are
fully complementary to narrow-band work:  whereas narrow-band imaging
probes a thin shell of redshift space, deep spectroscopy (admittedly
over a smaller solid angle) probes a pencil beam of look-back time.
The resolution of low-dispersion optical spectrographs are also better
matched to typical \lya\ line widths than filters with widths of 3000
\kms.

% FIGURE 10

\begin{figure}[!ht]
\plotfiddle{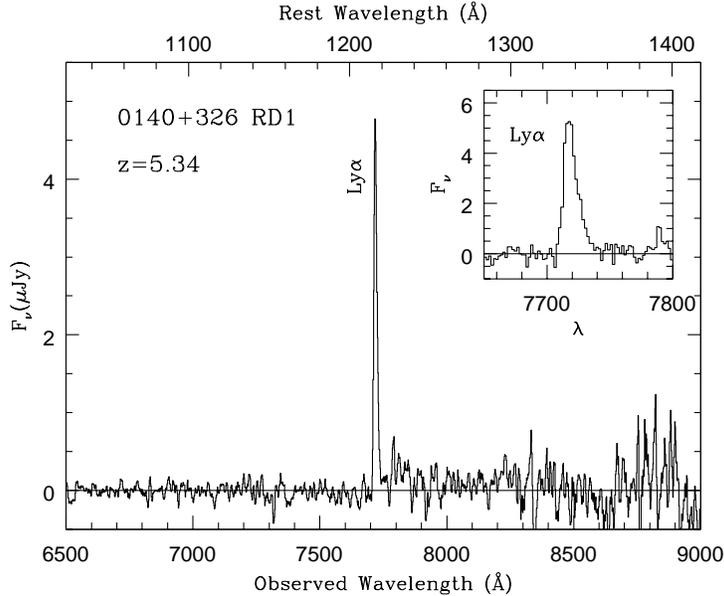}{2.8in}{0}{60}{60}{-180}{-150}

\caption[Spectrum of 0140+326~RD1 at $z = 5.34$]{Co-averaged spectrum
of the serendipitously discovered galaxy 0140+326~RD1 at $z = 5.34$ ($I
= 26.1$; Dey \etal 1998).  The total exposure time is 36.2~ksec.  Note
the strong \lya\ emission and the continuum discontinuity short-ward of
the line which confirms the redshift identification.  The `features'
observed in the continuum are largely due to residuals from the
subtraction of strong telluric OH emission lines (\eg 8344 \AA).  The
inset shows the asymmetric profile of the \lya\ emission line.  Figure
courtesy \markcite{Dey:98}Dey {et~al.} (1998).}

\label{rd1}
\end{figure}

At Berkeley, we are conducting a careful analysis of our deepest
archival spectroscopic exposures on the Keck telescopes obtained in
good meteorological conditions (see Fig.~\ref{serendips}).  In a 1.5
hour spectrogram at moderate-dispersion ($\lambda / \Delta \lambda
\simeq 1000$) with the LRIS camera, the limiting flux density probed
for spectrally unresolved line emission in a 1~arcsec$^2$ aperture is
$S_{\rm lim} (5 \sigma) \approx 1 \times 10^{-17} \ergcm2s$ at $\lambda
\approx 9300$ \AA\ \markcite{Stern:99b}(Stern {et~al.} 1999a).  The limit is strongly wavelength
dependent.  A single slitmask exposure typically covers $\approx 600$
square arcseconds, or one-sixth of a square arcminute.  In a year, up
to $\simgt 2$ square arcminutes can be covered, implying reasonable
statistics on the surface density of high-redshift line emitters, only
somewhat compromised when portions of the mask are dedicated to
photometric high-redshift candidates or when observations target rich
clusters.  This work is in progress \markcite{Manning:00}(Manning {et~al.} 2000) and the reader
will appreciate that distinguishing \lya\ serendips from lower-redshift
interlopers can be challenging and is best done with follow-up
spectroscopic and/or deep multi-band imaging observations.  Our
observations typically are biased towards redder wavelengths, which has
the unfortunate consequence that we are not sensitive to $z \sim 3$
serendips and thus cannot compare our results to well-documented
results of Steidel and collaborators at that redshift interval
\markcite{Steidel:96a, Steidel:96b, Dickinson:98, Steidel:99}(\eg Steidel {et~al.} 1996a, 1996b; Dickinson 1998; Steidel {et~al.} 1999).  We
are currently emphasizing serendipitous \lya-emitting galaxies at $4.5
\simlt z \simlt 5.5$.  We have four reasonable candidates in the
$\approx 1.6$ square arcminutes analyzed thus far, not including
candidates behind the Abell~2390 cluster.  One source is confirmed at
$z = 5.34$ \markcite{Dey:98}(Dey {et~al.} 1998).  We also have a few candidates at $z > 5.5$
which are not included in the current discussion.  The measured surface
density is $\approx 2 \pm 1$ candidate \lya\ emitters per square
arcminute at $z = 5$.  Their continua are estimated to be quite faint,
$I \simgt 26.5$ (AB).  Serendipitously-identified absorption-break
galaxies at high redshift are also possible, but difficult at the faint
levels we now probe.

\subsection{Narrow-band Spectroscopy }
\label{subsectz5nbspec}

An additional hybrid method, combining elements of both the narrow-band
imaging and serendipitous longslit search techniques, involves
obtaining many limited-wavelength spectra through a medium-width
(OH-avoidance) filter \markcite{Crampton:00}(\eg Crampton 2000).  With a uniform
placement of thirty 2\arcsec\ wide, 9\arcmin\ long parallel slitlets,
\markcite{Crampton:00}Crampton (2000), using the Canada-France-Hawaii Telescope atop
Mauna Kea, probes 9 square arcminutes of sky per mask observation with a
narrow-band filter plus low dispersion grism.  The survey is designed
to search for \lya\ at $6.38 < z < 6.64$, corresponding to a region of
both low OH emission and low H$_2$0 absorption.  This hybrid approach
has the advantage of providing spectra of all targets over a short
wavelength region.  This is important for distinguishing high-redshift
\lya, which can be identified both by line profile and by the presence
of a strong continuum decrement across the line.  Lower-redshift
emission features might also be identified from the presence of
secondary lines, such as [\ion{O}{3}]$\lambda$ 4959 \AA\ near
[\ion{O}{3}]$\lambda$ 5007 \AA, or the double profile of
[\ion{O}{2}]$\lambda\lambda$ 3727,3729 \AA.  Finally, as mentioned in
\S 4.4, the higher resolution of spectroscopy relative to narrow-band
imaging is more sensitive to the typical emission line widths, making
this hybrid approach an efficient method for probing the distant
Universe.

\subsection{Targeted Searches }

Finally, we briefly mention some targeted programs to identify
high-redshift galaxies associated with distant quasars.  A handful of
galaxies within a few arcseconds of distant quasars have been
identified at redshifts corresponding to damped \lya\ absorptions in
the quasar spectra \markcite{Djorgovski:96}(\eg Djorgovski {et~al.} 1996).  Searching for
galactic companions to distant quasars has also proved successful.
\markcite{Djorgovski:85}Djorgovski {et~al.} (1985) report on an emission-line-selected source with a
probable redshift of $z = 3.218$ associated with the quasar
PKS~1614$-$051 at $z = 3.209$.  More recent programs have used
combinations of narrow-band imaging and photometric color selection to
search for quasar companions at $z > 4$.  Several teams have identified
companions to BR~1202$-$0725 at $z = 4.695$ \markcite{Djorgovski:95,
Hu:96, Petitjean:96}(Djorgovski 1995; Hu {et~al.} 1996; Petitjean {et~al.} 1996).  \markcite{Hu:96b}Hu \& McMahon (1996) report on two companions to
BR~2237$-$0607 at $z = 4.55$.  \markcite{Djorgovski:00b}Djorgovski {et~al.} (2000) presents a status
report of an on-going program to search for protoclusters around
$\approx 20$ quasars at $z > 4$.  Preliminary results imply co-moving
number densities of protogalaxies two to four orders of magnitude
higher than expected for the field.  These results support biased
galaxy formation models, in which galaxies preferentially form in the
densest peaks of the primordial density field.

\subsection{Summary }

In Table~\ref{tabsurfden} we summarize the surface densities measured
(estimated) from the various search techniques described above.  We see
that the non-optical search techniques have very low surface densities,
most spectacularly for GRBs whose surface density is $\simlt 10$ per
year per Hubble volume.  The optical/near-IR techniques which are
sensitive to normal, star-forming galaxies, typically have surface
densities of a few per square arcminute per unit redshift.  In
\S\ref{sectz5bias} we consider the biases in the different techniques
and discuss how independent the methods may be.

% TABLE 5
\begin{table}[ht!]
\caption{Surface Densities of High-Redshift Galaxies}
\footnotesize
\begin{center}
\begin{tabular}{lcclc}
\hline\hline
Technique &
$z$ range &
\# / sq. arcmin &
Limiting Flux Density &
Reference \\
\hline
radio sources & all $z$ & $\approx 0.03$ & $S_{\rm 1.4GHz} > 1$ mJy & 1 \\
sub-mm galaxies & all $z$ & $\approx 0.09$ & $S_{\rm 850 \mu m} > 2$ mJy & 2 \\
X-ray sources & all $z$ & $0.27 \pm 0.04$ & $10^{-15} \ergcm2s$ & 3 \\
GRBs & all $z$ & $\approx 10^{-7}$ / yr & \nodata & \nodata \\
Lyman-break galaxies & $z \sim 3$ & $0.68 \pm 0.02$ & $R^{\rm AB} \leq 25.0$ & 4 \\
Lyman-break galaxies & $z \sim 4$ & $0.21 \pm 0.02$ & $I^{\rm AB} \leq 25.0$ & 4 \\
photometric candidates & $4.5 - 5.5$ & $\approx 1.0$ / unit-$z$ & $I_{814}^{\rm AB} \leq 26.5$ & 5 \\
narrow-band surveys, CADIS  & $z \sim 5.75$ & 0.2 / unit-$z$ & $5.0 \times 10^{-17} \ergcm2s$ (5$\sigma$) & 6 \\
narrow-band surveys, Hawaii & $z \sim 4.6$ & $\approx 3$ / unit-$z$ & $1.5 \times 10^{-17} \ergcm2s$ (5$\sigma$) & 7 \\
grism survey, NICMOS & $0.7 - 1.9$ & 0.5 & $2.0 \times 10^{-17} \ergcm2s$ & 8 \\
serendip survey, Berkeley & $4.5 - 5.5$ & $\approx 2 \pm 1$ / unit-$z$ & $1.0 \times 10^{-17} \ergcm2s$ (5$\sigma$) & 9 \\
\hline
\end{tabular}
\end{center}
\medskip

\emph{Notes.---} References:  (1) \markcite{Becker:95}Becker {et~al.} (1995); (2)
\markcite{Hughes:98}Hughes {et~al.} (1998); (3) \markcite{Hasinger:98}Hasinger {et~al.} (1998); (4) \markcite{Steidel:99}Steidel {et~al.} (1999); (5)
\markcite{FernandezSoto:99}Fern\`andez-Soto {et~al.} (1999); (6) \markcite{Thommes:98}Thommes {et~al.} (1998); (7)
\markcite{Cowie:98}Cowie \& Hu (1998).; (8) \markcite{McCarthy:99c}McCarthy {et~al.} (1999); (9) \markcite{Stern:99b}Stern {et~al.} (1999a).

\label{tabsurfden}
\end{table}
\normalsize

\section{``Protogalaxies'' at Low-Redshift}
\label{sectz5lowz}

The above discussion has focussed on identifying young galaxies at the
highest accessible redshifts.  The observed optical emission then
necessarily samples the rest-frame ultraviolet.  However, galaxy
formation is an ongoing process; locally we see several galaxies with
metallicities close to primordial \markcite{Kunth:94, Thuan:97a}(Kunth {et~al.} 1994; Thuan \& Izotov 1997).  These
nearby young systems sample a different segment of the galaxy
luminosity function from the high-redshift systems discussed above ---
the expectation is that high-redshift galaxies are the progenitors of
present-day massive galaxies ($L^*$ systems and larger) while the local
low-metallicity systems are all very low-mass dwarf galaxies.
Nevertheless, high-signal-to-noise ultraviolet observations of the
local protogalaxy population provide a very useful laboratory for
studying and understanding the high-redshift population.

As mentioned earlier, models predict strong ($50 - 200$ \AA\ equivalent
width) \lya\ emission from young, dust-free galaxies forming their
first generation of stars \markcite{Charlot:93}(\eg Charlot \& Fall 1993).  However, {\it
International Ultraviolet Explorer} ({\it IUE}) observations of local
star-forming galaxies revealed \lya\ strengths considerably weaker than
predicted by case B recombination:  the \lya/H$\beta$ intensity ratio
was always found to be $\simlt 10$, as opposed to the theoretical value
of 33.  Furthermore, some galaxies showed \lya\ absorbtion rather than
emission.  Small amounts of dust intermixed with the extended neutral
gas was the assumed culprit \markcite{Hartmann:84}(\eg Hartmann, Huchra, \& Geller 1984).
\markcite{Hartmann:88}Hartmann {et~al.} (1988) showed evidence for an anti-correlation of
\lya\ strength with metallicity, which conformed to a simplistic scheme
of chemical enhancement and motivated searches for strong
\lya\ emission in distant galaxies with anticipated primordial
abundances.  More recently, {\it HST} observations of the two most
metal-deficient galaxies known, I~Zw~18 \markcite{Kunth:94}($Z = Z_\odot /
51$; Kunth {et~al.} 1994) and SBS~0335$-$052 \markcite{Thuan:97a}($Z = Z_\odot /
40$; Thuan \& Izotov 1997) show \lya\ in {\em absorption} rather than emission,
at odds with the results of \markcite{Hartmann:88}Hartmann {et~al.} (1988).

\markcite{Kunth:98a}Kunth {et~al.} (1998) present {\it HST} ultraviolet spectra of eight \ion{H}{2}
galaxies covering a wide range of metallicity.  The observations were
designed to cover both the \lya\ region and the region around \ion{O}{1}
$\lambda 1302$ \AA\ and \ion{Si}{2} $\lambda 1304$ \AA.  The former
region allows study of the \lya\ emission and absorption properties and
an estimate of the \ion{H}{1} column, while the latter allows a crude
estimate of the chemical composition of the gas and a measure of the
velocity of the gas with respect to the systemic velocity of the system
as measured from optical emission lines.  Surprisingly, they find that
the primary indicator of \lya\ strength is kinematics, not metallicity.
The four systems with metallic lines static with respect to the ionized
gas show damped \lya\ absorption, while the four systems with \lya\
emission show the metallic lines blueshifted by $\approx 200$ \kms\
with respect to the ionized gas.  The implications are that even nearly
primordial clouds undergoing star formation have sufficient dust columns
to suppress \lya\ emission provided the kinematics of the neutral gas
allows resonant scattering of the \lya\ emission.  In all cases reporting
\lya\ emission in the \markcite{Kunth:98a}Kunth {et~al.} (1998) sample, an asymmetric profile
with a sharp blue cutoff is observed. 

In addition to the local star-forming galaxies with (1) broad, damped
\lya\ absorption centered at the wavelength corresponding to the
redshift of the \ion{H}{2} gas and (2) galaxies with \lya\ emission
marked by blueshifted absorption features, \markcite{Kunth:98b}Kunth {et~al.} (1999) notes a
third morphology of \lya\ line that is occasionally observed in the
local Universe:  (3) galaxies showing `pure' \lya\ emission, \ie
galaxies which show no \lya\ absorption whatsoever and have symmetric
\lya\ emission profiles.  \markcite{Terlevich:93}Terlevich {et~al.} (1993) presents {\it IUE}
spectra of two examples:  C0840+1201 and T1247$-$232, both of which are
extremely low-metallicity \ion{H}{2} galaxies.  \markcite{Thuan:97b}Thuan, Izotov, \& Lipovetsky (1997)
present a high signal-to-noise ratio {\it HST} spectrum of the latter
galaxy, noting that with $Z = Z_\odot / 23$, it is the lowest
metallicity local star-forming galaxy showing \lya\ in emission.  At
higher signal-to-noise ratio and higher dispersion than the {\it IUE}
spectrum, the line shows multiple absorption features near the redshift
of the emission, bringing into question the `pure' designation.
\markcite{TenorioTagle:99}Tenorio-Tagle {et~al.} (1999) have proposed a scenario to explain the variety
of \lya\ profiles based on the hydrodynamical evolution of superbubbles
powered by massive starbursts.

This scenario and observations of local star-forming galaxies have
two important implications for studies of high-redshift protogalaxies.
First, they provide a natural explanation for asymmetric profiles which
seem to characterize high-redshift \lya\ (\eg see Fig.~\ref{rd1}), but
also imply that although the asymmetric profile may be a sufficient
condition for identification of a strong line with \lya, it is not a
necessary condition.  This point is particularly important for judging
the identification of serendipitous and narrow-band survey emission
sources whose spectra are dominated by a solitary, high equivalent width
emission line.  Second, if \lya\ emission is primarily a function of
kinematics and perhaps evolutionary phase of a starburst, attempts to
derive the co-moving star-formation rate at high redshifts from \lya\
emission will require substantial and uncertain assumptions regarding the
relation of observed \lya\ properties to the intrinsic star formation
rate; use of the UV continuum ($\lambda \approx 1500$ \AA) may be
preferred for measuring star formation rates.

\section{Biases in Distant Galaxy Search Techniques}
\label{sectz5bias}

We now consider the biases which affect the various galaxy search
techniques discussed in \S\ref{sectz5notopt} and \S\ref{sectz5optir}.
Each technique provides a shaded view of the deep extragalactic
Universe; we enumerate these differences in the same order that the
various techniques were presented.

Radio galaxies are classical AGN.  This implies, according to the
prevailing models of active galaxies, the presence of a supermassive
black hole.  Nearby radio galaxies are associated with large early-type
galaxies, both cDs in the centers of clusters and giant elliptical
galaxies.  Conventional wisdom states that HzRGs are the precursors
to these early-type systems; our bias may simply be that HzRGs sample
the most massive systems.  At the least, though,
HzRGs necessitate the presence of a supermassive black hole in the
early Universe.  The time scale of supermassive black hole formation
is not well-known \markcite{Loeb:93}(\eg Loeb 1993).  Are black holes the seeds
of early galaxy formation, or are they the fruit of it?  The radio source
itself most likely affects the assembly of the galactic sub-units,
triggering star-formation along the radio jets in at least some cases
\markcite{Dey:97}(\eg Dey {et~al.} 1997).  If HzRGs are indeed found to be the precursors
of the most massive early-type systems, radio galaxies at high redshift
may prove to be most important for studies of the formation of galaxy
clusters and large scale structure.

Sub-mm and infrared-selected (LIRG) sources require much of their
bolometric luminosity to be dust-reprocessed radiation emitted at long
wavelength.  In an extreme scenario, one could imagine a
dust-enshrouded starburst in which no radiation short-ward of a few
microns escaped a restricted spatial region.  Such a source at high
redshift would escape detection in contemporary optical/near-infrared
surveys; it would perhaps only be detected in the 850$\mu$m sub-mm
region.  Have any such sources been identified already?  Astrometric
uncertainties and the large beam-size of the SCUBA detector leave the
question open.  It is conceivable that one or more of the
\markcite{Hughes:98}Hughes {et~al.} (1998) sub-mm sources in the HDF lack optical/near-infrared
identifications, even to the extremely deep limits of the HDF.  Though
studies of field sub-mm sources are in their nascent phases, it is
likely these systems represent those galaxies in the distant Universe
undergoing the most vigorous bursts of star formation.  We are far from
the time when we can reliably relate these distant sources to the local
census of galaxies, but conceiving of the sub-mm population as the more
distant cousins of ultraluminous infrared galaxies (ULIRGs) seems
plausible, if not likely \markcite{Lilly:99}(\eg Lilly {et~al.} 1999).  At the least, sub-mm
galaxies are important as they produce a significant fraction ($\geq
15$\%) of the total bolometric output of the Universe averaged over all
wavelengths and epochs \markcite{Lilly:99}(Lilly {et~al.} 1999) and largely (entirely?) account
for the sub-mm background.  Sub-mm galaxies may be the precursors of
metal-rich spheroids.

X-ray selection of distant sources likely necessitates the presence of
an AGN, implying that studies of the most distant X-ray sources may be
most important for understanding the formation of supermassive black
holes.  Extended cluster X-ray sources derive from the thermal
brehmsstrahlung in hot, ionized intercluster media.  Galaxy clusters
sample the deepest potential wells in the Universe on Mpc scales.
Studies of the most distant X-ray clusters will provide valuable
information on the formation of large-scale structure.  In particular,
several authors have noted the sharp dependence of the evolution of
cluster abundances on basic cosmological parameters
\markcite{Eke:96}(\eg Eke {et~al.} 1996).

The first optical identification of a GRB occurred less than two years ago,
and our understanding of these intriguing sources, though much improved,
is still rather sparse.  Are GRBs more prevalent in young stellar systems
with many massive remnants?  Are they correlated with SNe explosions?
It is premature to comment too deeply on what biases GRB-selected host
galaxies may yield on our understanding of the distant Universe.  At the
least, they are not {\em all} heavily reddened, which has implications
for the prevalence of dust in the Universe.

Lyman-break galaxies and photometrically-selected young galaxies at
high redshift are biased against dusty systems and older populations
which would redden the observed spectral energy distributions and
likely lead to redshift ambiguities.  The main bias is that these
techniques only search for galaxies containing a very young, OB
star-rich, stellar population which dominates the rest-frame continuum
in the wavelength range $\lambda\lambda 1216 - 1700$ \AA.  A small
contribution by an older population of stars will not be very
noticeable at the observed, deep-ultraviolet wavelengths available for
study at high redshift.   Near-infrared studies with spectrographs on
the new generation of $9 \pm 1$~m telescopes, and later from NGST, will
provide critical information on the masses, kinematics, ages, and dust
abundance in this population.  Are we only skimming the envelope of the
most dust-poor young systems?  The relation of the Lyman-break galaxies
to the local galaxy population remains uncertain.

\lya\ searches for distant galaxies have a clear, but uncertain,
observational bias; strong \lya\ emission ($W_\lambda > 20$ \AA) tends
to be biased against dusty galaxies, or at least against galaxies whose
neutral gas has unfavorable geometry/kinematics.  Spectroscopy of the
Lyman-break population reveals that approximately half of the galaxies
show \lya\ in {\em absorption}.  In fact, deep narrow-band imaging
compared to broad-band images can show these systems as negative holes
\markcite{Steidel:00}(Steidel 2000)!  However, most astronomers involved in
\lya\ surveys are restricting their search to emission-line sources.
The surface densities measured (see Table~\ref{tabsurfden}) are
comparable to those measured for Lyman-break galaxies.  The implication
is that a large population of strong \lya\ emitters exist which have
fainter continua than the limits of the photometric surveys.  If the
morphology of the \lya\ emission primarily depends upon the kinematics
of the neutral gas as suggested by ultraviolet studies of local
metal-poor galaxies (\S\ref{sectz5lowz}), then caution should be
advised in deriving star formation rates and densities from the
\lya-emitting galaxies.

\section{Results:  First Glimpses towards the Dark Ages}
\label{sectz5results}

Observational study of normal galaxies at $z \simgt 3$ is still a
rather recent phenomenon.  The last invited review on the subject to
this journal, only five years past \markcite{Pritchet:94}(Pritchet 1994), began with the
sad pronouncement that the predicted widespread population of primeval
galaxies had thus far escaped detection.  At the time, our knowledge of
individual sources at $z > 3$ was restricted to a dozen or so radio
galaxies and a larger census of quasars.  Now the pioneering work of
Steidel and collaborators has uncovered several hundreds of Lyman-break
systems at $z \simgt 3$ and spectroscopically confirmed galaxies have
recently crossed the $z = 5$ barrier.  Studies of the earliest epochs
of the Universe are no longer restricted to AGN and quasar
enthusiasts:  since 1997 December \markcite{Franx:97}(Franx {et~al.} 1997) the most distant
sources known to astronomers have consistently been apparently normal,
star-forming galaxies.  Though this review emphasizes the search
techniques for identifying distant galaxies, we now briefly consider
what has been and can be learned about the early Universe from these
studies.

% FIGURE 11
\begin{figure}[!t]
\plotfiddle{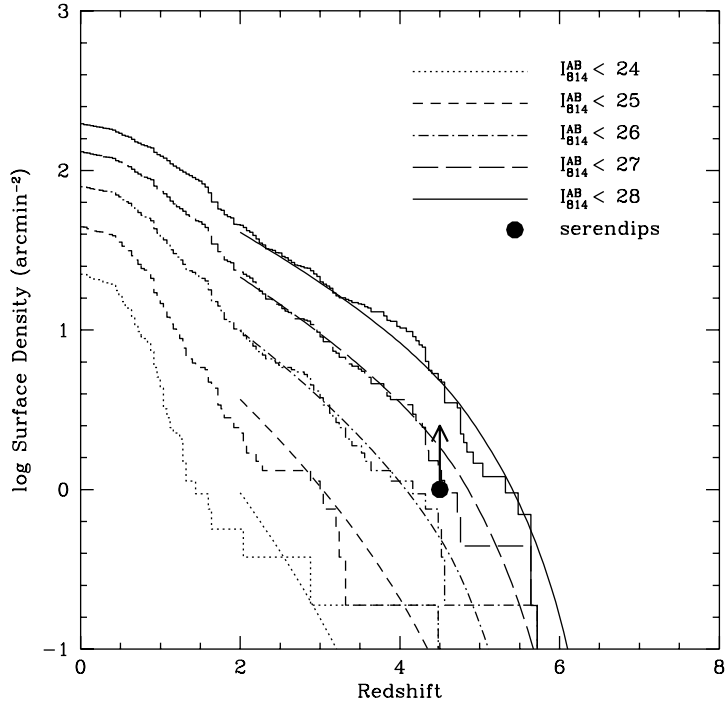}{3.5in}{0}{50}{50}{-150}{-80}

\caption[Cumulative surface density of galaxies]{Cumulative surface
density of galaxies as a function of redshift and limiting
$I_{814}^{\rm AB}$ magnitude.  Histograms are Stony Brook photometric
redshift measurements of the cumulative surface density of galaxies in
the (Northern) HDF.  Curves are fits to the histograms assuming a
parametrization of the evolving galaxy luminosity function (see text).
The point at $z = 4.5$ indicates our estimate of the lower limit to the
observed surface density of high-redshift serendipitous \lya-emitters
($S_{\rm Ly\alpha} \simeq$ 1.0 $\times
10^{-17} \ergcm2s$, 5 $\sigma$).  Figure
courtesy \markcite{Lanzetta:00}Lanzetta (2000).}

\label{surfden}
\end{figure}

\subsection{Surface Densities of High-Redshift Galaxies}

Our various types of distant galaxies may be illustrative of peculiar,
rare denizens of the cosmic past (the few powerful steep-spectrum radio
sources at $z > 4$) or they may be more modest star-forming citizens,
representative of an era when normal galaxies were modest fractions of
$L^*$ and $M^*$, typical luminosities and masses of the current cosmic
epoch near $z = 0$.

In Table~\ref{tabsurfden} we summarized the measured (estimated)
surface densities of the various classes of distant sources discussed.
Of particular note are the optical/near-infrared selected sources which
present our best hope at being identified with precursors of
present-day typical galaxies.

Fig.~\ref{surfden} illustrates the cumulative galaxy surface density as
a function of redshift and $I_{814}^{\rm AB}$ magnitude --- \ie the
surface density of galaxies of redshift greater than a given redshift
for different magnitude thresholds.  For example, to a limiting
magnitude of $I_{814}^{\rm AB} = 26$, $\approx 5$ galaxies per square
arcminute are expected at $z > 3$.  The histograms derive from Stony
Brook photometric redshift measurements in the (Northern) HDF.  The
curves are fits to the measurements, assuming that the evolving galaxy
luminosity function is parameterized by the \markcite{Schechter:76}Schechter (1976)
luminosity function, \begin{equation} \phi(L)dL = \phi^* (L/L^*)^\alpha
\exp(-L/L^*) d(L/L^*), \end{equation} where $\phi(L)$ is the number
density of galaxies per unit comoving volume and the characteristic
galaxy luminosity at rest-frame 1500 \AA\ $L_{*,1500}$ evolves with
redshift as \begin{equation} L_{*,1500} = L_{*,1500}^{z = 3} (1 +
z)^\beta.  \end{equation} \markcite{Lanzetta:00}Lanzetta (2000) finds that the
photometric redshifts are best fit with a characteristic absolute
magnitude at 1500 \AA\ $M_{*,1500}^{z = 3} = -19.5 \pm 0.5$, a
moderately steep luminosity function $\alpha = -1.43 \pm 0.05$, and an
evolving characteristic luminosity where $\beta = -0.7 \pm 0.4$.
Also plotted in Fig.~\ref{surfden} is our estimated lower limit to the
surface density of serendipitous sources at $z > 4.5$ from longslit
searches ($> 1$ arcmin$^{-2}$).

Substantial caveats regarding this plot should be kept in mind before
extragalactic astronomers halt spectroscopic redshift surveys.  First,
the modest angular size of the HDF implies that substantial cosmic
variance may skew the results, particularly at the brightest absolute
magnitudes for each redshift interval.  \markcite{Gwyn:00}Gwyn (2000) notes
significant differences in the distribution of photometric redshifts
between the HDF-N and the HDF-S.  Similarly, since the HDF samples a
very small volume of the local Universe, the low-redshift surface
densities are poorly determined.  Second, the redshifts are based on
photometric determinations rather than spectroscopic determinations.
As shown in \S\ref{subsectz5photz}, photometric measurements are fairly
robust at the magnitude and redshift ranges tested thus far.  However,
photometric redshifts are poorly tested in the difficult spectroscopic
redshift ranges of $1 \simlt z \simlt  2.5$ and $z \simgt 4$, as well
as at the extremely faint flux limits plotted in Fig.~\ref{surfden}.
Finally, these surface densities are based on optically-selected
objects.  If a large population of dust-enshrouded galaxies exists,
they may be missed in the optical surveys.

\subsection{Star Formation Rates and the Cosmic Star Formation History}
\label{subsectz5sfhist}

If the ionization is dominated by hot, young stars, the observed flux
of the \lya\ emission line may be used to estimate a lower bound to the
star formation rate $\dot{M}$ in a galaxy.  Using the case B
recombination \lya/H$\alpha$ ratio of $\approx 10$
\markcite{Osterbrock:89}(Osterbrock 1989), and the \markcite{Kennicutt:83}Kennicutt (1983) conversion from
H$\alpha$ luminosity to $\dot{M}$, \markcite{Madau:98}Madau {et~al.} (1998) find $\dot{M} \sim
0.7 \times 10^{-42}~h_{50}^{-2}~L_{{\rm Ly}\alpha}~M_\odot~{\rm
yr}^{-1}$ where $L_{{\rm Ly}\alpha}$ is measured in units of ergs~s$^{-1}$
($q_0 = 0.5$; $\dot{M}$ is $\approx 3.3$ times larger for $q_0 =
0.1$).  The star formation rate may also be estimated from the observed
ultraviolet continuum emission.  \markcite{Madau:98}Madau {et~al.} (1998) calculate that a
population older than 100~Myr will have $\dot{M} \approx
10^{-40}~L_{1500}~M_\odot~{\rm yr}^{-1}$ where $L_{1500}$, the
luminosity at 1500 \AA, is measured in units of ergs~s$^{-1}$\AA$^{-1}$.
\markcite{Leitherer:95}Leitherer, Carmelle, \&  Heckman (1995) models yield similar results for a different IMF
and ages less than 10~Myr.  These are lower limits to $\dot{M}$ since
no correction to the ultraviolet flux for either internal absorption or
dust extinction has been made (see \S\ref{subsectz5dust}).

With sufficient numbers of well-observed distant systems, we may
begin studying the star formation history of the Universe as a
function of co-moving volume.  The Canada-France Redshift Survey
\markcite{Lilly:95}(CFRS; Lilly {et~al.} 1995) demonstrated strong luminosity evolution
in the blue field galaxy population between $z = 0$ and $z = 1$.
\markcite{Madau:96}Madau {et~al.} (1996), using the early results of photometric selection in
the HDF, integrated the results at $z < 5$ into a coherent picture of the
star formation history of the Universe and suggested that the global star
formation peaks between $z = 1$ and $z = 2$, in remarkable agreement with
predictions based on the co-moving \ion{H}{1} density traced by \lya\
absorption systems \markcite{Pei:95}(Pei \& Fall 1995), with hierarchical models in a cold
dark matter-dominated Universe \markcite{Baugh:98}(Baugh {et~al.} 1998), and with the quasar
luminosity function \markcite{Cavaliere:98}(Cavaliere \& Vittorini 1998).

Substantial caveats temper this result, however:  (1) the number of
spectroscopically measured redshifts between $z = 1$ and $z = 2$ is
small.  \markcite{Connolly:97}Connolly {et~al.} (1997) use optical and near-infrared data to
estimate photometric redshifts in the HDF at $1 < z < 2$ in order to
span this redshift ``desert'' for which few strong spectroscopic
features are shifted into the optical regime.  Until a substantial
number of confirmed redshifts have been obtained, however, photometric
redshifts in the $1 < z < 2$ range are not well constrained.  Thus the
epoch thought to be the most productive in terms of star formation is
also the least well measured.  (2) The small area ($\approx$ five
square arcminutes) covered by the HDF makes global parameters inferred
from it vulnerable to perturbations from large scale structure.  (3)
The existence of the peak at $z \sim 1.5$ is contingent upon the
completeness of the estimates of global star formation at higher
redshift.  The $z \sim 4$ point in the analysis of \markcite{Madau:96}Madau {et~al.} (1996) was
based on a single object in the HDF at $z = 4.02$ \markcite{Dickinson:98}(Dickinson 1998)
and should thus be treated as an uncertain lower limit.  More recent
measurements, based on larger-area surveys, show that the star
formation density is significantly higher at $z \simgt 4$
\markcite{Steidel:99}(Steidel {et~al.} 1999).  Finally, (4) both the photometric and the
\lya\ search techniques require the objects to be UV-bright.  It is
possible that a substantial fraction of star-forming activity in dusty
high-redshift systems has been overlooked thus far.  The recent
discoveries of a relatively large number of resolved sources in small
area sub-mm surveys performed by SCUBA suggest this to be the case
\markcite{Hughes:98, Barger:98, Dey:99a}(\eg Hughes {et~al.} 1998; Barger {et~al.} 1998; Dey {et~al.} 1999).

%We now address the star formation density at $z \sim 5$.  At the end of
%\S\ref{sectz5optir} we determined a conservative total surface density
%of $\approx 2 \pm 1$ serendipitous galaxies per unit z per square
%arcminute ($4.5 < z < 5.5$) based solely on serendipitous emission line
%sources and not including photometrically-selected absorption-break
%sources.  For this simplistic calculation, we assume that the
%serendipitous sources at $z \sim 5$ have a characteristic star
%formation rate of 12 $M_\odot$ yr$^{-1}$, equal to $L^*$ for the
%Lyman-break population at $z \sim 3$ for negligible dust extinction
%\markcite{Dickinson:98}(Dickinson 1998).  This is close to the mean of the recent $z > 5$
%spectroscopic measures \markcite{Dey:98, Weymann:98, Spinrad:98}(Dey {et~al.} 1998; Weymann {et~al.} 1998; Spinrad {et~al.} 1998).
%\markcite{Steidel:99}Steidel {et~al.} (1999) has shown that the luminosity function of
%Lyman-break galaxies at $z \simeq 3$ and $z \simeq 4$ are
%well-described by the Schechter function with $\alpha = -1.60$;  we
%assume the same form is appropriate at $z \sim 5$.  Following
%\markcite{Steidel:99}Steidel {et~al.} (1999), we integrate the luminosity function to $0.1 L^*$
%to calculate the cosmic star-formation rate per comoving volume at $z
%\simeq 5$.  The results of this calculation is presented in
%Fig.~\ref{starformden}.

% FIGURE 12

\begin{figure}[!t]
\plotfiddle{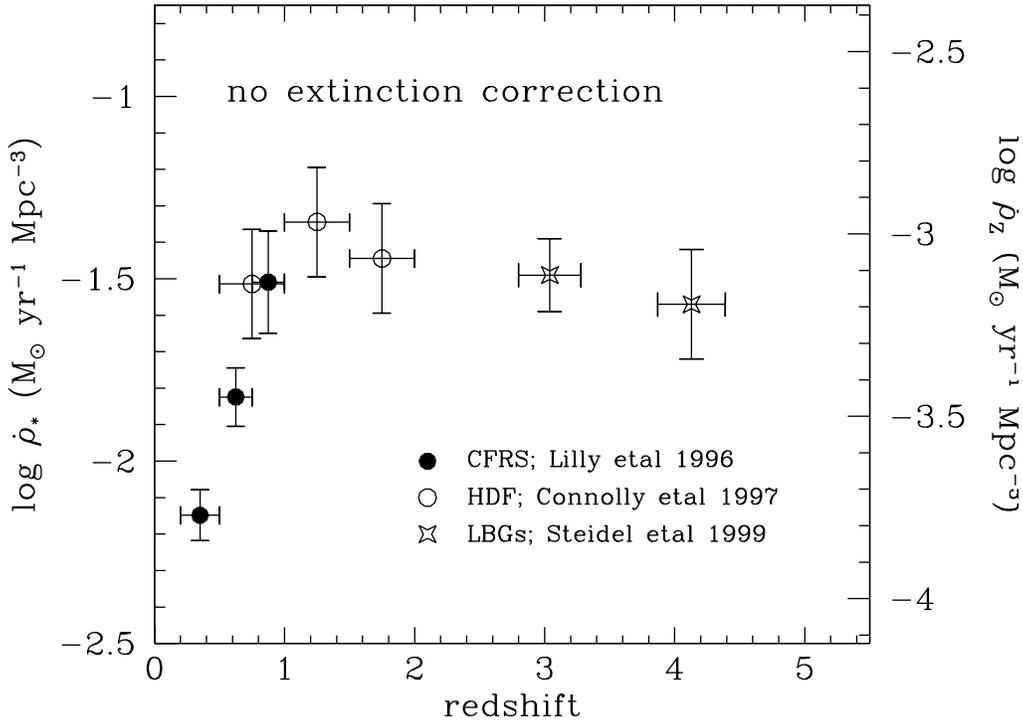}{3.6in}{-90}{50}{50}{-200}{300}

\caption[Star formation history of the Universe]{The star formation
history of the Universe.  Points at $z < 1$ are from the Canada France
Redshift Survey \markcite{Lilly:96}(CFRS; Lilly {et~al.} 1996); points at $1 \simlt z \simlt
2$ are from near-infrared studies of the HDF \markcite{Connolly:97}(Connolly {et~al.} 1997); and
points at $z \simeq 3$ and $z \simeq 4$ are from studies of Lyman-break
galaxies \markcite{Steidel:99}(LBGs; Steidel {et~al.} 1999).  For a Salpeter IMF, the cosmic
metal ejection density, $\dot{\rho}_z$, is 1/42 of the cosmic star
formation density, $\dot{\rho}_*$ \markcite{Madau:96}(Madau {et~al.} 1996).}

%The point at $z = 5$ is from our estimate of the surface density of
%(serendipitously-identified) \lya-emitting galaxies at $4.5 < z < 5.5$
%(see text for details).

\label{starformden}
\end{figure}

Fig.~\ref{starformden} illustrates a recent determination of the star
formation history of the Universe.  In sharp contrast to the initial
results of \markcite{Madau:96}Madau {et~al.} (1996), the UV luminosity density of the Universe
is approximately constant at $2 \simlt z \simgt 4$.  Preliminary
results from the Berkeley longslit searches suggest that it remains
constant to $z \simeq 5.5$.  The evolution of the star formation
density of the Universe, as probed by apparently normal, star-forming
systems, implies that their evolution follows a significantly different
trajectory than that of AGN, especially quasars.  Both radio-quiet and
radio-loud quasars show a considerable density decrease beyond $z \sim
3$ \markcite{Dunlop:90, Hook:98, Cavaliere:98}(\cf Dunlop \& Peacock 1990; Hook \& McMahon 1998; Cavaliere \& Vittorini 1998), falling by a factor
of $\sim 3$ from $z = 3$ to $z = 4$.

\subsection{Effects of the IGM:  Hints of the Gunn-Peterson Effect?}

\begin{figure}[!t]
\plotfiddle{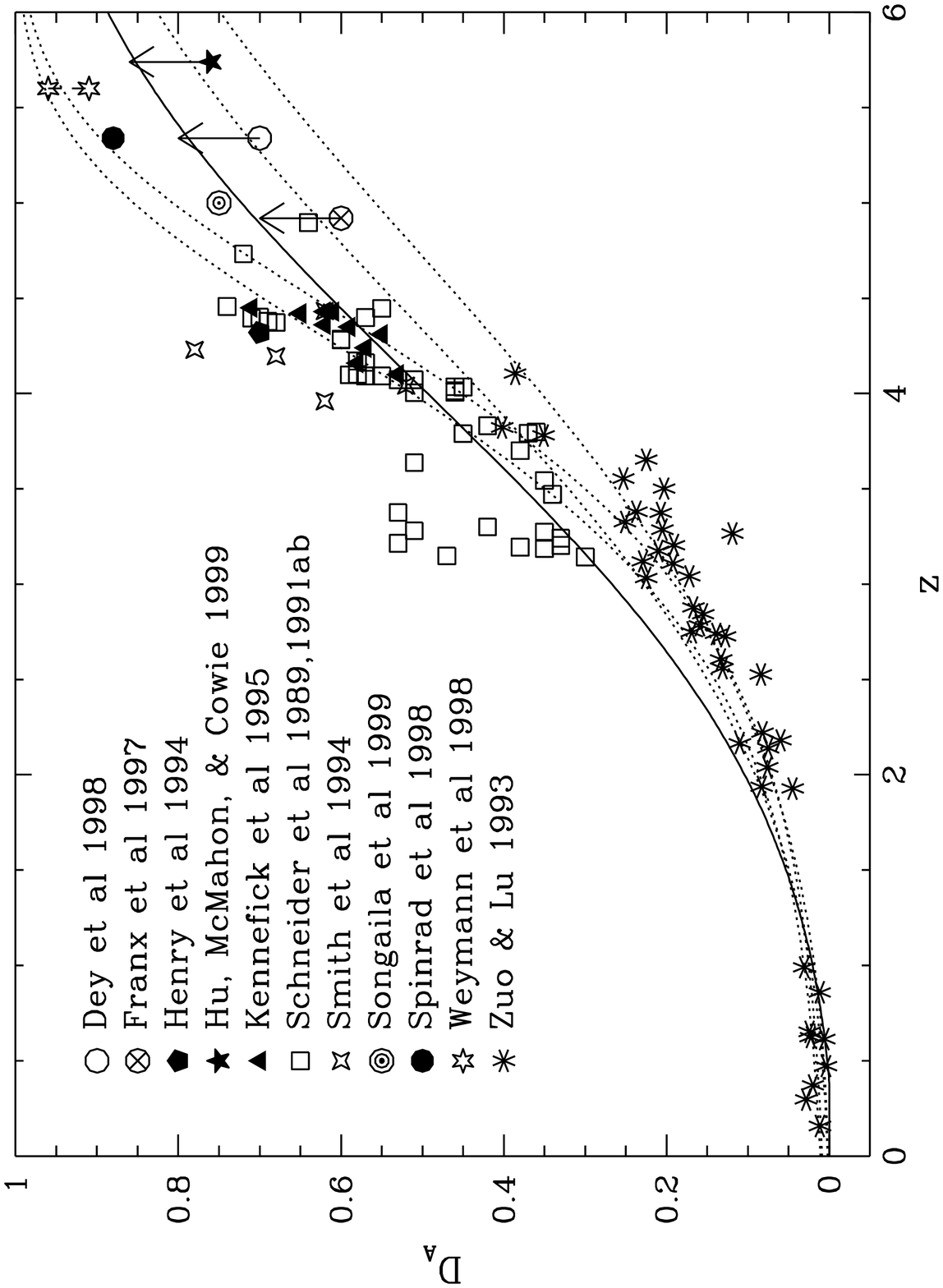}{3.6in}{-90}{50}{50}{-200}{300}

\caption[$D_A$ as a function of redshift]{Values of the continuum
depression blue-ward of \lya\ \markcite{Oke:82}($D_A$;  Oke \& Koryanski 1982) plotted as a
function of redshift from several observational samples \markcite{Dey:98,
Franx:97, Henry:94, Hu:99, Kennefick:95, Schneider:89, Schneider:91a,
Schneider:91b, Smith:94, Songaila:99, Spinrad:98, Weymann:98, Zuo:93}(Dey {et~al.} 1998; Franx {et~al.} 1997; Henry {et~al.} 1994; Hu {et~al.} 1999; Kennefick {et~al.} 1995; Schneider, Schmidt, \& Gunn 1989; Schneider, Schmidt, \&  Gunn 1991a, 1991b; Smith {et~al.} 1994; Songaila {et~al.} 1999; Spinrad {et~al.} 1998; Weymann {et~al.} 1998; Zuo \& Lu 1993).
The \markcite{Madau:95}Madau (1995) model (solid line) and several parametrizations
from \markcite{Zhang:97}Zhang {et~al.} (1997) (dotted lines) are also plotted (see text).
\markcite{Zuo:93}Zuo \& Lu (1993) may be systematically offset from the other
measurements.  At $z \simgt 5$ are we perhaps seeing the hints of
hydrogen absorption in excess of what is expected from the \lya\ forest
alone?  Could this be the first indications of the long-anticipated
Gunn-Peterson effect?}

\label{da}
\end{figure}

The ability of astronomers to calculate reliable photometric redshifts
at $z > 3$ is largely due to the neutral hydrogen absorption in the
intergalactic medium (IGM).  Absorption below the redshifted Lyman
limit, Ly$\beta$, and \lya\ strongly modulates the observed optical
spectra of the objects we are interested in.  At sufficiently high
redshifts, the continuum depression blue-ward of \lya\ (the
\lya\ forest) dominates over those decrements associated with higher
members of the Lyman series.  \markcite{Oke:82}Oke \& Koryanski (1982) define the $D_A$ parameter
to measure the strength of the \lya\ decrement \begin{equation} D_A
\equiv 1 - {{f_\nu(\lambda\lambda 1050 - 1170)_{\rm obs}}\over
{f_\nu(\lambda\lambda 1050 - 1170)_{\rm pred}}} \end{equation} (see
Fig.~\ref{da}).  The question is: at what redshift does the
\lya\ forest become an impenetrable ``jungle''?

% FIGURE 13

This is an intriguing question, as we are now observing galaxies at
early-enough cosmic epoch that we may perhaps encounter the edge of the
``Dark Era'' --- prior to the complete reionization of the IGM when the
Universe is still optically thick below the rest wavelength of \lya.
Theoretical estimates suggest that the dark era ends between $z = 5$
and $z = 20$ \markcite{Shapiro:95, Rees:95, MiraldaEscude:97, Madau:98}(Shapiro 1995; Rees 1995; Miralda-Escude \& Rees 1997; Madau {et~al.} 1998).
Observationally, we would expect a (Gunn-Peterson) trough of complete
absorption for some interval short-ward of \lya, corresponding to the
epoch when the Universe was completely opaque to \lya\ photons.

How can we measure the Gunn-Peterson effect?  For objects of $I \sim
26$, the obvious problem is one of signal-to-noise ratio.  Combining
the results of our spectrophotometry with deep {\it HST} images, we
measure rather large values of $D_A$ for HDF~3-951.0 \markcite{Dey:98}($z = 5.34$;
$D_A = 0.88$; Dey {et~al.} 1998) and HDF~4-473.0 \markcite{Weymann:98}($z = 5.60$; $D_A = 0.91
- 0.96$; Weymann {et~al.} 1998).  Our concern about the break amplitude arises
from its strength:  the \markcite{Madau:95}Madau (1995) theoretical estimate of the
contribution of the \lya\ forest to $D_A$ (solid line in Fig.~\ref{da})
is only $\approx 0.79$ at $z = 5.34$ and $\approx 0.83$ at $z = 5.60$.
This extrapolation assumes a distribution of high and low optical depth
foreground \lya\ clouds causing Lyman series absorption in the spectrum
of a distant quasar or galaxy.  The dotted lines in Fig.~\ref{da}
represent different models of $D_A$ from \markcite{Zhang:97}Zhang {et~al.} (1997).
Using the data of \markcite{Steidel:87}Steidel \& Sargent (1987) and \markcite{Zuo:93}Zuo \& Lu (1993), they estimate
the mean intergalactic \lya\ opacity $\tau_\alpha$ as a function of
redshift using two parametrizations of $\tau_\alpha$:  $\tau_\alpha = A
(1 + z)^{3.46}$ and $\tau_\alpha = A e^{\beta (1 + z)}$.  The scatter
around the theoretical curves is substantial, even at lower redshifts,
so the high values of $D_A$ at $z > 5$ may simply reflect the usual
scatter observed in that parameter.  However, the possibility to
directly measure the epoch of reionization is not to be overlooked.
The best chance for this will be with the discovery of a bright(er)
object at $z > 5$, either in the form of a quasar, or a magnified
galaxy behind a rich galaxy cluster.

\subsection{Morphology}

At $z = 5$, the Universe is only 890~$h_{50}^{-1}$ Myr old,
corresponding to a lookback time of 93.2\%\ of the age of the
Universe.  Any galaxies observed at these early epochs must necessarily
be in their youth, and any information we can obtain on them is of the
utmost interest to studies of galaxy formation.  In particular, if
these galaxies are truly primeval objects forming their first
generation of stars, their morphologies can provide constraints on
galaxy formation models.  If the formation of a galactic spheroid
occurs via the monolithic collapse of a protogalactic cloud
\markcite{Eggen:62}(\eg Eggen {et~al.} 1962), then the bulk of its star formation might
occur in a small region kiloparsecs in extent;  such high-redshift
protogalaxies will appear as compact, luminous objects \markcite{Lin:92}(Lin \& Murray 1992).
Alternatively, if galaxy formation is a hierarchical process
\markcite{Baron:87, Baugh:98}(\eg Baron \& White 1987; Baugh {et~al.} 1998), protogalaxies may appear as a
multitude of unresolved subgalactic clumps embedded in a more diffuse
gaseous halo.

\begin{figure}[!t]
\plotfiddle{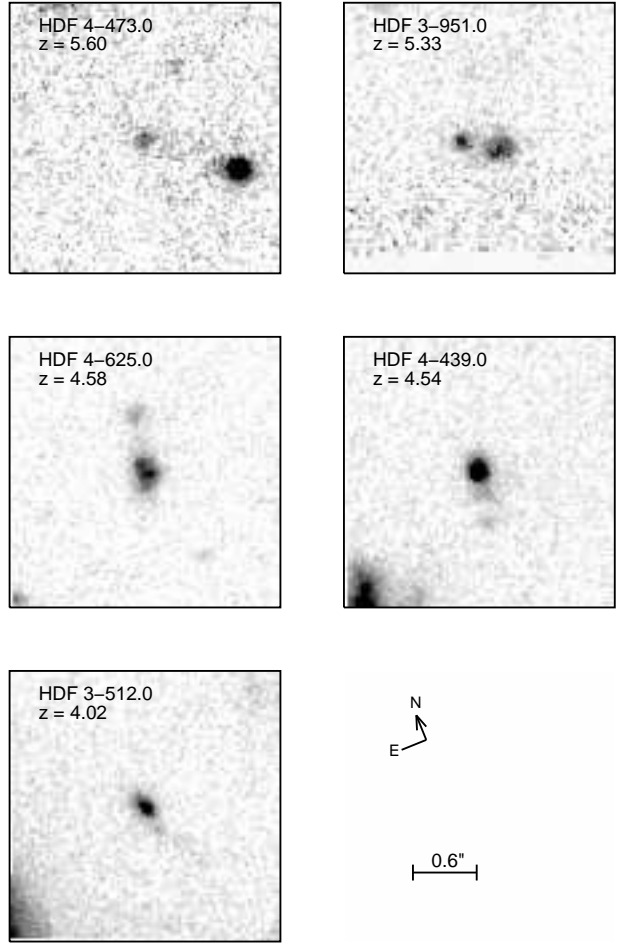}{5.5in}{0}{50}{50}{-130}{10}

\caption[Montage of galaxies at $z > 4$ in the HDF]{Montage of
spectroscopically-confirmed $z > 4$ galaxies in the HDF, from the
drizzled F814W ($I_{814}$) images.  Note that most of the galaxies are
compact, but show evidence of merging activity (interactions), in the
form of multiple nuclei, multiple components, and/or extended,
low-surface brightness tidal tails.}

\label{hdf3_951image}
\end{figure}

All of the confirmed $z > 4$ galaxies in the HDF
(Fig.~\ref{hdf3_951image}) have compact, but resolved morphologies,
with deconvolved half-light radii of $\approx 0\farcs2$ ($\approx 1 -
2~ h_{50}^{-1}~$ kpc), comparable to the values found for many of the
$z \sim 3$ Lyman-break galaxies \markcite{Giavalisco:96}(Giavalisco {et~al.} 1996).  The sizes are
clearly sub-galactic, suggestive of the hierarchical scenarios of
galaxy formation.  HDF~3-951.1, the brighter component of HDF~3-951.0
($z = 5.33$ for both), contains substructure with a second ``hot spot''
approximately 0\farcs12 east of the core, at a projected separation of
0.66 $h_{50}^{-1}$ kpc.  We speculate that this is either a knot of
star formation (bright in the rest-frame ultraviolet), or evidence of
multiple nuclei.  The projected proximity of HDF~3-951.2 adds weight to
the hypothesis that this is a dynamically-bound system, and that we are
witnessing a merger event.  HDF~4-439.0 at $z = 4.54$ and HDF~4-625.0
at $z = 4.58$ also both have multiple components.  Lyman-break galaxies
at $z \sim 3$ often exhibit either disrupted morphologies or multiple
components \markcite{Giavalisco:96, Steidel:96b, Bunker:99}(\eg Giavalisco {et~al.} 1996; Steidel {et~al.} 1996b; Bunker, Moustakas, \& Davis 1999).
However, HDF~4-473.0 at $z = 5.60$ shows no evidence of substructure.

% FIGURE 14

\subsection{The Role of Dust}
\label{subsectz5dust}

Since most of our observations sample restframe ultraviolet wavelengths,
derived quantities such as star formation rates are sensitive to gas and
dust extinction.  As noted earlier, \lya\ is particularly vulnerable since
it is a resonance line whose escape is dependent upon the distribution
of neutral gas.

One method of estimating the amount of reddening, used by
\markcite{Meurer:97, Pettini:99, Dickinson:98, Bunker:99, Steidel:99}Meurer {et~al.} (1997); Pettini {et~al.} (1999a); Dickinson (1998); Bunker {et~al.} (1999); Steidel {et~al.} (1999), is
to assume a very young ($\approx 10^7$ yr age) starburst and ascribe any
color excess in the emitted ultraviolet to extinction.  The uncertainty,
of course, is that aging of a starburst will also cause reddening.
Typical extinction corrections are factors of $2 - 7$ near 1500 \AA,
depending upon the extinction law applied.  \markcite{Meurer:97}Meurer {et~al.} (1997) derive a
correction factor of $\approx 15$, in part due to different assumptions
regarding the spectral slope of the underlying, unreddened population.

Near-infrared photometry offers the possibility of more reliable estimates
of reddening, since at these wavelengths photometry is insensitive to the
details and ages of the hottest stars.  The ground-based (Keck/NIRC)
$J$-band detection of 0140+326~RD1 by \markcite{Armus:98}Armus {et~al.} (1998) suggests
substantial reddening ($A_V > 0.5$, for \markcite{Bruzual:93}Bruzual \& Charlot (1993) models
with ages less than $10^8$ yr), which is somewhat surprising given its
non-detection in deep sub-mm observations, strong \lya\ emission, and
similar star formation rates inferred from its rest-frame ultraviolet
continuum and \lya\ flux density.

{\it HST}/NICMOS imaging offers deep, reliable near-infrared photometry
and will be a valuable asset for measuring reddening in Lyman-break
galaxies.  For example, \markcite{Weymann:98}Weymann {et~al.} (1998) use NICMOS F110W and F160W
photometry to limit the reddening and star formation rate of HDF~4-473.0
($z = 5.60$).  They find that the spectral energy distribution limits
the reddening to $0.00 \simlt E(B-V) \simlt 0.12$ and the star formation
rate to $8~M_\odot~{\rm yr}^{-1} \simlt \dot{M} \simlt 19~M_\odot~{\rm
yr}^{-1}$.  This modest amount of extinction is consistent with the
rather strong \lya\ emission line emerging from this distant galaxy.

Near-infrared spectroscopy offers a potent tool for studying the dust
content, mass, age, and kinematics of distant galaxies.
\markcite{Pettini:98}Pettini {et~al.} (1998) recently reported a pilot program of near-infrared
spectra of the well-studied, rest-frame optical, nebular emission lines
from \ion{H}{2} regions in five Lyman-break galaxies at $z \simeq 3$.
The observations used the CGS4 spectrometer on the United Kingdom
Infrared Telescope and targeted the redshifted Balmer and [\ion{O}{3}]
emission lines.  H$\beta$ luminosities, uncorrected for intrinsic dust
extinction, imply star formation rates of $20 - 270~ h_{70}^{-2}~
M_\odot~{\rm yr}^{-1}$ ($q_0 = 0.1$); that is, typically a factor of
several larger than that inferred from the UV continuum of these
galaxies.  The implication is that an extinction of $1 - 2$ magnitudes
at 1500 \AA\ may be typical of the Lyman-break population.  Velocity
dispersions of $\sigma \simeq 70~$\kms were reported in four out of the
five galaxies, suggesting virial masses $M_{\rm vir} \approx (1 - 5)~
\times~10^{10}~M_\odot$.  The relative redshifts of \lya\ emission,
interstellar absorption, and nebular emission lines vary by several
hundred \kms, suggestive of large-scale outflows.   Similar scale
outflows are common in regions of rapid star formation locally.

Future higher-resolution (spectral and spatial) observations of these and
related rest-frame optical transitions redshifted into the near-infrared
should better address the kinematics and possibly the light-element
abundances of young, forming galaxies.

\subsection{Galaxy Clustering at High Redshift}
\label{subsectz5lss}

Astronomers generally assume that the galaxy distribution traces the
underlying matter distribution; comparisons of the large-scale
distribution of galaxies at early cosmic epoch therefore is a
potentially powerful means of discriminating cosmology and mechanisms of
structure formation \markcite{White:97}(\eg White 1997).  A simple constant of
proportionality known as the ``bias parameter'', $b$, relating galaxy
and mass fluctuations, is the conventional parametrization:
$\delta_{\rm gal} = b \delta_{\rm mass}$.  In principle, many physical
mechanisms could lead to a relation of this form \markcite{Dekel:87}(\eg Dekel \& Rees 1987),
though gravitational instability in a massive dark matter halo is
currently popular as numerical simulations and semi-analytic models of
this ``dark halo'' model are consistent with observations
\markcite{Baugh:98, Mo:99}(\eg Baugh {et~al.} 1998; Mo, Mao, \& White 1999).

Sufficient numbers of galaxies at $z \simgt 3$ have now been reliably
identified that observational measurements of their spatial
distribution is now possible.  \markcite{Giavalisco:98}Giavalisco {et~al.} (1998) reports on the
angular clustering of Lyman-break galaxies at $z \sim 3$.  The slope of
a power-law parametrization of the angular correlation function,
$w(\theta) = A_w\theta^{-\beta}$, is $\beta \sim
0.9$, similar to galaxy samples in the local and intermediate-redshift
Universe.  Applying the Limber transform to $w(\theta)$ yields the
comoving spatial correlation length.   \markcite{Giavalisco:98}Giavalisco {et~al.} (1998) report
$r_0 = 4.2^{+0.9}_{-1.0}~h_{50}^{-1}$ Mpc ($q_0 = 0.5$) at the median
redshift of their Lyman-break survey, $\bar{z} = 3.04$.  The value is
similar to that of local spiral galaxies and approximately half that of
local early-type galaxies; it is comparable or slightly larger than
comoving spatial correlation lengths determined for
intermediate-redshift galaxies.  The strong clustering is broadly
consistent with biased galaxy formation theories, suggesting that the
Lyman-break systems are associated with massive dark matter halos.

\markcite{Steidel:98}Steidel {et~al.} (1998) report a large structure of Lyman-break galaxies at
$z \simeq 3.09$ in the SSA22 field which they interpret in the context
of cold dark matter cosmological models.  Dark halo models of galaxy
formation predict that galaxies of a given mass should form first in
regions of the highest density and that these regions should be
strongly clustered spatially.  \markcite{Adelberger:98}Adelberger {et~al.} (1998) measure the bias
parameter, $b$ for the \markcite{Steidel:98}Steidel {et~al.} (1998) $z \simeq 3.09$ structure:
considering 268 Lyman-break galaxies in six 9\arcmin\ $\times$
9\arcmin\ fields with spectroscopic redshifts at $z \simeq 3$,
\markcite{Adelberger:98}Adelberger {et~al.} (1998) perform a counts-in-cell analysis, measuring the
fluctuations in galaxy counts in cells of differing comoving volume.
They find that the variance in cubes of comoving side length $(15.4,
23.8, 22.8)~h_{50}^{-1}$~Mpc is $\sigma^2_{\rm gal} \sim 1.3 \pm 0.4$.
Following the methodology of \markcite{Peacock:94}Peacock \& Dodds (1994), the implied bias
factor is $b = (6.0 \pm 1.1, 1.9 \pm 0.4, 4.0 \pm 0.7)$ for these
spatial scales.  The result is broadly consistent with simple dark halo
models of structure formation, in which matter fluctuations are
Gaussian, have a linear power-spectrum shape similar to that determined
locally ($\Gamma \sim 0.2$), and Lyman-break galaxy luminosities are
correlated with their mass.  The results are largely independent of
cosmology.  \markcite{Adelberger:98}Adelberger {et~al.} (1998) note that measurements of the
Lyman-break galaxy masses could, in principal, distinguish cosmological
scenarios.

\subsection{A Brief Comparison to Theories of Galaxy Formation}
\label{subsectz5thy}

The \markcite{Pritchet:94}Pritchet (1994) review on primeval galaxies speculates that
progenitors of galaxies like our Milky Way should be very numerous,
regardless of appearance.  The data, in the form of images and spectra
of distant $z \simgt 3$ Lyman-break systems and faint number counts,
currently suggests that galaxies start as many smaller subclumps and
halos so that the local density of $L^*$ galaxies ($0.015~h_{50}^3~{\rm
Mpc}^{-3}$) may substantially underestimate the co-moving space density
of small objects at $z = 5$.

The dominant paradigm for understanding the Lyman-break population is
the ``dark halo'' model:  galaxies quiescently form stars at the bottom
of the potential wells of massive dark matter halos.  Assuming
rest-frame UV luminosities correlate with galaxy mass, the brightest
Lyman-break galaxies should form first in regions where the density is
highest.  Since these regions are expected to be strongly clustered
spatially, the high-redshift, large-scale structures discussed in
\S\ref{subsectz5lss} are explained naturally.  Over time, the halos
merge, forming the massive galaxies we see locally.

\markcite{Baugh:98}Baugh {et~al.} (1998) present a semianalytic model of this hierarchical
galaxy formation scenario, focusing on the properties of Lyman-break
galaxies at $z \simeq 3$.  With a ``suitable'' choice of parameters, they
are able to reproduce the observed Lyman-break galaxy properties for
cold dark matter (CDM) cosmologies with both $\Omega_0 = 1$ and
$\Omega_0 < 1$.  At high redshift, galaxies have very small bulges or
no bulge at all:  typical half-light radii are $\sim 1~ h_{50}^{-1}$~
kpc, in good agreement with the $z \sim 3$ results of
\markcite{Giavalisco:96}Giavalisco {et~al.} (1996) and the HDF images at $z > 4$.  \markcite{Baugh:98}Baugh {et~al.} (1998)
also reproduce the strongly biased spatial distribution, with $b \simeq
4$ and a comoving correlation length $r_0 \simeq 8~ h_{50}^{-1}$~Mpc at
$z \simeq 3$.  These models predict that the average $L^*$ galaxy today
was in $\approx 4$ sub-units at $z = 1$ and $\approx 14$ sub-units at
$z = 5$.

However, the \markcite{Baugh:98}Baugh {et~al.} (1998) models fair less well with respect to
star formation rates.  They predict that at $z \simeq 3$, most galaxies
are only forming a few solar masses of stars per year and only a very
small fraction have star formation rates in excess of $40~ h_{50}^{-2}~
M_\odot~ {\rm yr}^{-1}$.  This is at odds with more recent estimates of
the rest-UV extinction of the $z \simeq 3$ Lyman-break population, \eg
the near-infrared spectroscopic results of \markcite{Pettini:98}Pettini {et~al.} (1998).  The
hierarchical models also predict that galaxies form the bulk of their
stars at relatively low redshift \markcite{Baron:87}(\eg Baron \& White 1987), with $\simeq
50$\%\ of the stars formed since $z \simeq 1$.  The \markcite{Baugh:98}Baugh {et~al.} (1998)
models predict that cosmic star formation history peaks around $z = 1 -
2$, in rough concordance with the first measurements of the comoving
star formation history by \markcite{Madau:96}Madau {et~al.} (1996).  More recent measurements,
discussed in \S\ref{subsectz5sfhist}, still show the rapid evolution in
comoving star formation rate from $z = 0$ to $z \sim 1.5$, but, with
larger samples of high-redshift objects less vulnerable to cosmic
variance and improved consideration of rest-frame UV extinction, the
revised plots show a plateau in the comoving star formation density
from $z \sim 1.5$ to $z \sim 4$ (Fig.~\ref{starformden}).

An alternate view of the Lyman-break population maintains that these
galaxies are primarily collision-induced galactic starbursts, triggered
by small, gas-rich satellite galaxies \markcite{Lowenthal:97,
Somerville:99, Kolatt:99}(Lowenthal {et~al.} 1997; Somerville, Primack, \&  Faber 1999; Kolatt {et~al.} 1999).  Using semianalytic models,
\markcite{Somerville:99}Somerville {et~al.} (1999) study the properties of individual galaxies in
the ``quiescent'' dark halo scenario, similar to that addressed by
\markcite{Baugh:98}Baugh {et~al.} (1998), in comparison to the ``collisional starburst''
scenario.  They argue that the high star-formation rates, small nebular
emission line widths \markcite{Pettini:98}($\sim 70~$ \kms; Pettini {et~al.} 1998), young ages
\markcite{Sawicki:98}(median age $\sim 25$~Myr; Sawicki \& Yee 1998), and high surface
densities are all better explained by the collisional starburst
model.  More recently, \markcite{Kolatt:99}Kolatt {et~al.} (1999) use high-resolution N-body
simulations to address the clustering properties of Lyman-break sources
in the collisional starburst model.  They find that although most
sources are relatively low mass in this scenario, they cluster around
high-mass halos and therefore exhibit the observed strongly biased
clustering.

\section{Conclusions and Thoughts for the New Millennium}
\label{sectz5conclude}

Progress in the field of distant galaxies has been rapid.  Young,
star-forming galaxies have been located by several means, and studied
from space and the ground to magnitudes as faint as $V = 28$ and
redshifts as large as $z = 5.7$, perhaps to even $z = 6.7$ or higher.
A population of faint, dusty galaxies has been detected in the sub-mm
region; their redshift distribution remains uncertain for now.

Observations with {\it HST} have established the prevalence of 
small protogalaxy candidates at high redshift.  How and when do
they merge to form the familiar Hubble types observed locally?

The last review of the subject of primeval galaxies in this journal,
only five years ago, was largely a census of non-detection limits and a
discussion of theoretical expectations of the predicted wide-spread
population of young galaxies at high redshift \markcite{Pritchet:94}(Pritchet 1994).  The
field is expanding so fast now that this review will be largely
outdated at the time of press.  We conclude with a brief discussion of
ripe avenues for the field of deep extragalactic studies.
Several of the next steps will involve new instruments and satellites
slated for commissioning in the coming months and years.  We also
suggest some ventures to longer wavelengths that those typically
employed in contemporary early Universe studies.  It is, of course, the
redshift that pushes us in that direction.

\begin{itemize}

\item{Currently, much of the early Universe extragalactic research
has focussed on simply procuring redshifts, identifying objects at
earlier and earlier cosmic epoch, and studying the spatial clustering
of (the Lyman-break) population.  Spectroscopy has the potential
to provide considerably more information.  Detailed, higher-resolution,
higher signal-to-noise ratio observations of some of the brighter
Lyman-break galaxies can probe the ages, kinematics, and abundances
of young (proto-)galaxies.  \markcite{Spinrad:99a}Spinrad {et~al.} (1999) suggests, from a
very limited sample, that \lya\ emission strength anticorrelates
with the strength of the rest-frame UV interstellar absorption
lines.  More recently, \markcite{Pettini:00}Pettini {et~al.} (1999b) have reported on intermediate
resolution, high signal-to-noise ratio spectroscopy of the lensed
galaxy MS~1512$-$cB58 at $z = 2.727$.  This detailed study probes
the stellar initial mass function (IMF) at early cosmic epoch, finding
no evidence for a flatter IMF (at the high-mass end) or an IMF
deficient in high-mass stars.  They measure a metallicity of $\approx
0.25$ solar and bulk outward motions of 200 \kms, which may be an
important mechanism of distributing metals in the IGM.  This pioneering
study lays the groundwork for follow-up studies on larger samples of
(unlensed) sources.}

\item{The new generation of near-infrared spectrographs on
large-aperture, ground-based telescopes such as the Keck Near Infrared
Spectrometer (NIRSPEC) and the Very Large Telescope (VLT) Infrared
Spectrometer And Array Camera (ISAAC), opens the window on observing
the early Universe to unprecedented redshifts.  The first
near-infrared detection of \lya\ emission will be a technological
feat.  What search techniques will prove most efficient at identifying
sources at $z \approx 10$, assuming that sufficiently luminous sources
have collapsed at that redshift?  These cameras should be sensitive to
line emission at flux densities $\sim 10^{-17} \ergcm2s$ out to $\sim
2.5 \mu$m in the windows between telluric OH emission.  An hour
spectrum with a 10~m telescope at resolving power $R \sim 2000$ should
yield a detection at a signal-to-noise ratio $S/N \sim 5$ for an
unresolved emission line of that strength.  A more-realistic,
slightly-resolved (spatially and spectrally) emission line would
require several hours to yield a $S/N$ of a few per resolution
element.  Detection of the anticipated $0.5 \mu$Jy continuum longward
of \lya\ will remain challenging:  NIRSPEC, as an example, will require
$\approx$ six hours on integration to reach $S/N \sim 1$ per resolution
element; multiple-binning will be requisite.}

\item{Several new telescopes, cameras, and satellite missions are
expected to be commissioned shortly.  At sub-mm wavelengths,
more-sensitive cameras with improved spatial resolution such as SCUBA+
and BOLOCAM will be having first light in the next few years.
Satellites such as {\it SIRTF}, {\it Chandra}, and {\it XMM} will open
up the mid-infrared and X-ray universe considerably.}

\item{With the prospect of {\it NGST} becoming more realistic, one can
consider deep imaging and spectroscopy in the infrared, perhaps to
$\sim 4 \mu$m.  This would open up the ``dark age'' to extreme,
unprecedented redshifts; at $z = 25$, \lya\ propogates to $3.16 \mu$m
and we are probing the Universe at a time less than 100~$h_{50}^{-1}~$
Myr after the Big Bang.  Imaging above and below this wavelength might
be an excellent diagnostic for the very early quasars, galaxies, and
supernovae.}

\item{The mm-region of the spectrum has been a good region for
molecular studies of our Milky Way galaxy.  In the future it may also
be the domain of choice for redshift coolant lines such as
[\ion{C}{2}]$\lambda 158 \mu$m.  This transition is potentially strong
in both \ion{H}{1} and \ion{H}{2} regions; some local starbursts emit
between $10^{-2}$ and $10^{-3}$ of the infrared (dust) continuum in
this line.  Its application to large redshifts is still uncommon, but
unless the $C/H$ ratio in young galaxies is disastrously low, it seems
worth attempting detection of high redshift [\ion{C}{2}]$\lambda 158
\mu$m with modern mm-interferometers.  At $z = 7$, for example, the
line redshifts to 1.26~mm.  The opportunity then would be substantial
-- even for some physical study of a very distant (gas-rich) stellar
system.}

\end{itemize}

\acknowledgements

We are indebted to our close collaborators Andrew Bunker and Arjun Dey
for extensive conversations on the subject of distant galaxies and
helping shape our conception of the deep Universe.  We also gratefully
acknowledge conversations and communications with Kurt Adelberger, Len
Cowie, Carlos De~Breuck, Mark Dickinson, George Djorgovski, Peter
Eisenhardt, James Graham, Esther Hu, Ken Lanzetta, Curtis Manning, Pat
McCarthy, Ed Moran, Leonidas Moustakas, George Smoot, Adam Stanford,
Chuck Steidel, Eduard Thommes, Wil van~Breugel, and Rogier Windhorst.
We thank Alberto Fern\'andez-Soto for providing Fig.~\ref{specphotz},
Ken Lanzetta for providing Fig.~\ref{surfden}, Trinh Thuan for sharing
the {\it HST}/GHRS spectrum of T1214$-$277, and Dave Hollenbach for
conversations regarding the [\ion{C}{2}]$\lambda 158 \mu$m emission
line strength.  We are also indebted to Sam Maxie for considerable
typographical efforts on the early draft of this manuscript.  We
acknowledge NSF Grant AST~95$-$28536 for supporting some of the
extragalactic research presented herein.  Research by DS has been
supported by IGPP/LLNL grant 98$-$AP017.

%\bibliographystyle{apj}
%% \bibliography

\end{document}